\documentclass[]{jfm}
 \usepackage{amsmath}
\usepackage{amssymb}
\usepackage{color}
\usepackage{float}
\usepackage{dcolumn}
\usepackage{bm}
\usepackage{graphicx}
\usepackage{newtxtext}
\usepackage{newtxmath}
\usepackage{natbib}
\usepackage{hyperref}
\usepackage{booktabs}
\usepackage{setspace}

\hypersetup{
    colorlinks = true,
    urlcolor   = blue,
    citecolor  = blue,
    linkcolor  = blue,
}

\newcommand{\RomanNumeralCaps}[1]

\title{A theoretical model for compressible bubble dynamics considering phase transition and migration }

\author{A-Man Zhang\aff{1,2,3}
	\corresp{\email{zhangaman@hrbeu.edu.cn}},
	Shi-Min Li \aff{1,3},  
	Run-Ze Xu \aff{1}, 
	Shao-Cong Pei\aff{1},
	Shuai Li\aff{1,2,3},
	\and Yun-Long Liu\aff{1,2,3}}

\affiliation{\aff{1}College of Shipbuilding Engineering, Harbin Engineering University, Harbin, 150001, China
	\aff{2}Nanhai Institute of Harbin Engineering University, Sanya, 572024, China
	\aff{3}National Key Laboratory of Ship Structural Safety, Harbin Engineering University, Harbin, China, 150001
}
\begin{document}
\maketitle

\begin{abstract}

A novel theoretical model for bubble dynamics is established that simultaneously accounts for the liquid compressibility, phase transition, oscillation, migration, ambient flow field, etc. The bubble dynamics equations are presented in a unified and concise mathematical form with clear physical meanings and extensibility. The bubble oscillation equation can be simplified to the Keller-Miksis equation by neglecting the effects of phase transition and bubble migration.
The present theoretical model effectively captures the experimental results for bubbles generated in free fields, near free surfaces, adjacent to rigid walls, and in the vicinity of other bubbles. Based on the present theory, we explore the effect of the bubble content by changing the vapor proportion inside the cavitation bubble for an initial high-pressure bubble. It is found that the energy loss of the bubble shows a consistent increase with increasing Mach number and initial vapor proportion. However, the radiated pressure peak by the bubble at the collapse stage increases with the decreasing Mach number and increasing vapor proportion. The energy analyses of the bubble reveal that the presence of vapor inside the bubble not only directly contributes to the energy loss of the bubble through phase transition but also intensifies the bubble collapse, which leads to greater radiation of energy into the surrounding flow field due to the fluid compressibility. 

\end{abstract}


\section{Introduction}
\label{sec:intro}

From the eruption of submarine volcanoes and underwater explosion \citep{khwwk,lhfww19,RN398}
 to the snapping of pistol shrimps \citep{vshl00,lsv01}, from targeted drug delivery and ultrasonic lithotripsy \citep{Lokhandwalla2001,fpb07,Maeda2019} to ultrasonic cleaning \citep{Verhaagen2016,Jaekyoon2018,Landel2021}, 
bubble dynamics holds significant importance across various academic areas and practical applications. The behavior of oscillating bubbles involves a complex interplay of factors such as fluid compressibility, bubble migration, and mass and heat transfer \citep{fa80,Qianxi11,Brujan2022,RN1349}. Understanding these complex physical mechanisms not only advances fundamental knowledge but also drives innovations in technologies reliant on bubble phenomena.

The Rayleigh-Plesset (RP) equation \citep{r17,plesset49} stands as a classical framework widely utilized to predict the oscillation behavior of spherical cavitation bubbles. While rooted in the assumption of incompressible fluids, it has provided foundational insights into various aspects of bubble dynamics, including nonlinear bubble oscillations \citep{h70,RN272,Storey2001,RN1356}, and linear interactions between multiple bubbles \citep{RN272,Harkin2001,baol06}. Over the years, researchers have developed numerous compressible models to address the limitations of the RP equation \citep{Herring1941,g52,kk56,pp86}. Examples include the Keller-Miksis equation \citep{kk56,km80}, known for its robust theoretical foundation,
and \citet{Chahine2018} incorporated the influence of bubble migration by integrating an incompressible migration term. \citet{gh02} employed the doubly asymptotic approximation (DAA) approach to develop equations that capture bubble oscillation and migration within a compressible flow environment for underwater explosion bubbles, and \citet{unified2023} derived the oscillation and migration equations in a compressible flow field under various environmental conditions based on the wave equation. However, it's difficult for their models to calculate the bubble dynamics in which the bubble contents are composed of both condensable and non-condensable gases.

Recent research by \citet{Zhong2020} and \citet{Rui2023} have revealed that, in addition to fluid compressibility and viscosity, the condensation and evaporation processes of vapor inside laser-induced and spark bubbles also significantly affect the dynamic characteristics of bubbles, particularly concerning the issue of energy loss after the second cycle of bubble oscillation. Furthermore, previous compressible bubble models using the adiabatic gas equation of state have struggled to accurately reproduce the energy loss during the multi-cycle oscillation of bubbles, regardless of how the initial conditions are configured \citep{Qingyun2018,cerbus2022experimental,PhysRevLett.132.104004}. This indirectly suggests that these compressible bubble models lack certain crucial physical mechanisms. In fact, the phase transition model of bubbles has been extensively studied in the previous studies, including the state equation of gases \citep{Gallo2023,Abbondanza2023}, the rate of phase transition \citep{Fuster2010,Yasuibook18}, and the temperature boundary layer near the bubble surface \citep{fa80,Hauke2007,Tian2022}. Moreover, previous studies predominantly analyzed the bubble migration under the assumption of incompressible fluids \citep{h70,RN272,Seo2010}, overlooking the impact of fluid compressibility. In this study, we will specifically examine the roles of phase transition and bubble migration in formulating a comprehensive bubble oscillation equation within the compressible fluid domain, 
and a new migration equation that accounts for the effects of fluid compressibility and condensation/evaporation will be deduced. 


Furthermore, quantitative analyses of bubble content remain challenging due to the complex mixture of water vapor and non-condensable gases within bubbles. To address this challenge, we manipulate the composition of bubble content by changing the initial vapor proportion inside the bubble at a constant initial internal bubble pressure, allowing for a systematic analysis of their influence on bubble dynamics. 
This study seeks to provide comprehensive insights into bubble dynamics, fostering advancements in both fundamental understanding and practical applications.

The structure of this paper is organized as follows. We firstly derive the theoretical model in detail in \S\ \ref{sec:rules_submission}, including the bubble oscillation equation, state equation of mixed gases, bubble migration equation, multiple bubble equation, and bubble equation of boundary effect. In \S\ \ref{3}, the theoretical model is fully validated by several bubble experiments in the free field, near boundaries, and near multiple bubbles. In \S\ \ref{4}, parametric studies on the effects of initial vapor proportion inner the bubble are conducted for an initially high-pressure bubble. Finally, this study is summarized and conclusions are made in \S\ \ref{5}.

\section{Theory}\label{sec:rules_submission}
\subsection{Bubble oscillation equation} 

The physical model of this study is characterized by a spherical bubble with a radius of $R$ oscillating in the compressible liquid. The bubble oscillation is coupled with phenomena such as bubble migration and phase transition. The fluid domain is treated as weakly compressible and satisfies the linear wave equation \citep{unified2023}. With the center of the bubble as the coordinate origin $o$, the wave equation in the spherical coordinate system $o - r \theta \phi$ is expressed as
 
\begin{equation}
\label{eq5}
\frac{1}{{{C}^{2}}}\frac{{{\partial }^{2}}\varphi }{\partial {{t}^{2}}}=\frac{1}{{{r}^{2}}}\frac{\partial }{\partial r}\left( {{r}^{2}}\frac{\partial \varphi }{\partial r} \right)+\frac{1}{{{r}^{2}}\sin \theta }\frac{\partial }{\partial \theta }\left( \sin \theta \frac{\partial \varphi }{\partial \theta } \right)+\frac{1}{{{r}^{2}}{{\sin }^{2}}\theta }\frac{{{\partial }^{2}}\varphi }{\partial {{\phi }^{2}}},
\end{equation}
{where $\varphi$ is velocity potential of liquids and $C$ is the sound speed.}

Assuming that the bubble keeps spherical oscillation, we define here that the bubble migrates along the direction of $\theta =0$. Thus, $\partial \varphi/\partial \phi=0$ and the third term at the right-hand side of Eq. (\ref{eq5}) vanishes because of the axisymmetry. The induced velocity potential $\varphi_{\rm f_{\rm s}}$ induced by a source at the origin with the strength of $f_{\rm s}(t)$ is 
\begin{equation} \varphi_{\rm f_{\rm s}}({\bm r},t) = -\frac{1}{{|\bm r|}} f_{\rm s}\left(t-\frac{|\bm r|}{C}\right).
\end{equation} 

According to the linearity characteristics of the wave equation, the superposition of a set of $\phi_{{\rm f}_{\rm s}}$ can produce the solution of the wave equation $\varphi_{\rm f}$ considering the source movement: 
\begin{equation}\label{eq-source-moving}
{\varphi_{\rm f}(\boldsymbol{r},t) = 
	-\frac{C}{\left(C-\boldsymbol{v}\cdot\boldsymbol{r}_{\rm t}/|\boldsymbol{r}_{\rm t}|\right)|\boldsymbol{r}_{\rm t}|} f\left(t-\frac{|\boldsymbol{r}_{\rm t}|}{C}\right)},
\end{equation}
where $\boldsymbol{r}_{\rm t}$ is the vector pointing from the source at $t-|\boldsymbol{r}_{\rm t}|/C$ to $\boldsymbol{r}$; $f$ is a function whose second-order derivative exists. The relative velocity vector $\boldsymbol{v}$ represents the velocity difference between the bubble migration velocity ${\bm v}_{\rm m}$ and the ambient flow velocity ${\bm u}_{\rm a}$. Once Eq. (\ref*{eq-source-moving}) is obtained, the velocity potential of the moving dipole $\varphi_{\rm q}$ can be calculated as
\begin{equation}\label{eq-dipole-moving}
\varphi_{\rm q} = \lim_{D \rightarrow0}\frac{1}{2D}\left(\varphi_{\rm f}(\boldsymbol{r}+\boldsymbol{e}D,t)-\varphi_{\rm f}(\boldsymbol{r}-\boldsymbol{e}D,t)\right),
\end{equation}
\noindent where the unit vector $\boldsymbol{e}$ indicates the direction along which the bubble migrates, and $D$ denotes the distance that is halfway between the point source and the sink.
%
%
%

Considering that the migration velocity is small relative to sound speed, the location variation of singularities could be ignored during the short time when the influences propagate from the singularities to the bubble surface. Thus, Eq. (\ref{eq-source-moving}) could be simplified, and the linear superposition of the velocity potential of the point source and dipole can be expressed as 

\begin{equation}
\label{eq-velocity-potential}
\varphi{(r,\theta,t)} = -\frac{1}{r}f{\left(t-\frac{r}{C}\right)} - \frac{\mathcal{\cos\theta}}{r^2}q{\left(t-\frac{r}{C}\right)} - 
\frac{1}{C}\frac{\cos\theta}{r}q'{\left(t-\frac{r}{C}\right)},
\end{equation} 
where $q$ is a function whose second-order derivative exists, and $q'$ represents the derivative of $q$ with respect to its argument.
 
The time derivative of $\varphi$ and the velocity of the bubble surface in the $r$ direction are 
 
\begin{equation}
\frac{\partial \varphi}{\partial t} =  -\frac{1}{r}f'{\left(t-\frac{r}{C}\right)} - \frac{\mathcal{\cos\theta}}{r^2}q'{\left(t-\frac{r}{C}\right)} - 
\frac{1}{C}\frac{\cos\theta}{r}q''{\left(t-\frac{r}{C}\right)}, 
\end{equation} 
and
\begin{equation}
\label{eq-phi-r}
u_{\rm r} = 
{{\left. \frac{\partial \varphi }{\partial r} \right|}_{r=R}} = 
\frac{f}{R^2}+ 
\frac{f'}{CR}  
+\frac{2\cos\theta}{R^3}q + 
\frac{2\cos\theta }{CR^2 }q',
\end{equation}
respectively; where $f'$ represents the derivative of $f$ with respect to its argument, and $q''$ denotes the second-order derivative of $q$ with respect to its argument. The terms of magnitude $1/C^2$ are ignored in Eq. (\ref{eq-phi-r}) and subsequent derivations. 
 
According to the continuity condition at the bubble surface considering the phase transition \citep{fa80}, the normal velocity of fluids at the bubble surface equals to $\dot R - \dot m/\rho$ ($\dot{R}$ denotes the time derivative of $R$; $\dot m$ is the net evaporation rate of mass per unit area of the bubble surface; $\rho$ is the liquid density). Then, the averaged kinetic boundary condition for the bubble oscillation can be expressed by



\begin{equation}
\label{kine}
 \int_{S_{\rm b}} u_{\rm r} \mathrm{d}S = 4\pi R^2 \left(\dot R- \frac{\dot m}{\rho}\right),
\end{equation}
where $S_{\rm b}$ denotes the area of the bubble surface.

Substituting Eq. (\ref{eq-phi-r}) into (\ref{kine}), $q$ and $q'$ are vanished such that

\begin{equation}
\label{kine1}
\frac{f}{R^2}+ 
\frac{f'}{CR} = \dot R- \frac{\dot m}{\rho}.
\end{equation}

Let $F(t) = f(t - R/C)$, then we have
\begin{equation}
\label{eq9}
\frac{\rm{d}\textit{F}}{\rm{d}\textit{t}}=\left( {1 - \frac{{\dot R}}{C}} \right)f'{|_{r = R}}.
\end{equation}

Taking the derivative of Eq. (\ref{kine1}) with respect to time and combining it with (\ref{eq9}) yields
 
 \begin{equation}
 \label{eq15}
 \left( {\frac{{C - \dot R}}{R} + \frac{{\rm{d}}}{{{{\rm{d}}t}}}} \right)\left( {\frac{R}{{C - \dot R}}\frac{{{\rm{d}}F}}{{{\rm{d}}t}}} \right) +\frac{\text{d}}{\text{d}t}\left( \frac{{{R}^{2}}\dot{m}}{\rho } \right) = 2R{\dot R^2} + {R^2}\ddot R,
 \end{equation}
where ${{{\rm{d}}F}}/{{{\rm{d}}t}}$ depends on different physical problems and environmental conditions. Once ${{{\rm{d}}F}}/{{{\rm{d}}t}}$ and $\dot m$ are determined, Eq. (\ref{eq15}) can be solved to obtain the bubble oscillation dynamics. 
As the effect of phase transition is neglected, the above equation is simplified to 

\begin{equation}
 \left( {\frac{{C - \dot R}}{R} + \frac{{\rm{d}}}{{{{\rm{d}}t}}}} \right)\left( {\frac{R}{{C - \dot R}}\frac{{{\rm{d}}F}}{{{\rm{d}}t}}} \right)  = 2R{\dot R^2} + {R^2}\ddot R .
\end{equation}

To obtain ${{{\rm{d}}F}}/{{{\rm{d}}t}}$, we apply the Bernoulli equation to the bubble surface in the moving coordinate system:

\begin{equation}
\label{eq-bernoulli}
{{\left. \frac{\partial \varphi }{\partial t} \right|}_{r=R}} - \boldsymbol{v} \cdot \boldsymbol{u}+ \frac{1}{2}{\left| \boldsymbol{u} \right|^2} + H = 0,
\end{equation}
\noindent where $\boldsymbol{v} \cdot \boldsymbol{u}$ can be expressed as the inner product of the vectors $\left( v\cos\theta, -v\sin\theta \right)$ and $\left( u_{\rm r}, u_\theta \right)$, with $v=|\bm v|$.
$H=\int_{{P_{\rm{a}}}}^{{P_{\rm{b}}}} {\rho^{-1}} {} \mathrm{d} p$ is the enthalpy difference at the bubble surface, and its zero-order term can be expressed as $(P_{\rm{b}}-P_{\rm{a}})/\rho$. Here, $P_\mathrm{a}$ represents the ambient pressure at the bubble center  
($P_{\rm{a}}$ includes the hydrostatic pressure at infinity $P_{\infty}$, the acoustic pressure, and the pressures induced by boundaries and other bubbles), and $P_\mathrm{b}$ denotes the liquid pressure at the bubble surface.

According to Eq. (\ref{eq-velocity-potential}), Eq. (\ref{eq-phi-r}) and (\ref{kine1}), the normal and tangential velocities of the bubble surface can be expressed as 

 \begin{equation}
\label{oscillav}
{{u}_{\rm r}}=\dot{R}-\frac{\dot{m}}{\rho }+\frac{2\cos \theta }{{{R}^{3}}}q+\frac{2\cos \theta }{C{{R}^{2}}}{q}' ,
\end{equation} 
and
\begin{equation}
u_{\rm \theta} = 
\frac{1}{R}
{{\left. \frac{\partial \varphi }{\partial \theta} \right|}_{r=R}} =
\frac{\mathcal{\sin\theta}}{R^3}q
+\frac{\sin\theta}{C R^2}q',
\end{equation}
respectively.

Integrating the Bernoulli equation over the bubble surface, all the terms containing $\theta$ could be eliminated. Consequently, both $q$ and $q'$ are eliminated and an equation containing only the unknown quantity $f'$ can be obtained such that

 \begin{equation}
\label{intergrate}
\begin{aligned}
&\int\limits_{S} {\left( {{{\left. {\frac{{\partial \varphi }}{{\partial t}}} \right|}_{r = R}} - {\bm v} \cdot {\bm u} + \frac{1}{2}{{\left| {\bm u} \right|}^2} + H} \right){\rm d}S} \\
&= \int\limits_0^\pi  {\left.{\left[ \begin{array}{l}
	- \frac{1}{r}f' - \frac{{\cos \theta }}{{{r^2}}}q' - \frac{1}{C}\frac{{\cos \theta }}{r}q'' - \left( {v\cos \theta , - v\sin \theta } \right) \cdot \left( {{u_{\rm r}},{u_{\rm \theta} }} \right)\\
	+ \frac{1}{2}\left( {u_{\rm r}^2 + u_\theta ^2} \right) + H
	\end{array} \right]}\right|_{r=R} 2\pi {R^2}\sin \theta {\rm d}\theta } \\
&= 4\pi {R^2}\left( { - \frac{{f'}}{R} + \frac{1}{2}{{\left( {\dot R - \frac{{\dot m}}{\rho }} \right)}^2} + \frac{1}{4}{v^2} + H} \right) = 0.
\end{aligned}
\end{equation}

Combine Eq. (\ref{eq9}) and (\ref{intergrate}), and the expression of ${{{\rm{d}}F}}/{{{\rm{d}}t}}$ in Eq. (\ref{eq15}) could be obtained:

 \begin{equation}
\label{dfdt}
\frac{\rm{d}\textit{F}}{\rm{d}\textit{t}}= R \left( {1 - \frac{{\dot R}}{C}} \right) \left(\frac{1}{2}{\left({{\dot R}-\frac{\dot m}{\rho}} \right)^2}+\frac{1}{4}{{v}^{2}}+H \right).
\end{equation}

Substituting the above expression into the Eq. (\ref{eq15}) and the bubble oscillation equation considering bubble migration and phase transition could be provided as

\begin{equation}
\label{unif}
\left( {\frac{{C - \dot R}}{R} + \frac{{\rm{d}}}{{{\rm{d}}t}}} \right)\left [{\frac{{{R^2}}}{C}\left( {\frac{1}{2}{\left({{\dot R}-\frac{\dot m}{\rho}} \right)^2} + \frac{1}{4}{{v}^2} + H} \right)} \right] +\frac{\text{d}}{\text{d}t}\left( \frac{{{R}^{2}}\dot{m}}{\rho } \right) = 2R\dot R_{}^2 + {R^2}\ddot R .
\end{equation}

Eq. (\ref{unif}) is the bubble oscillation equation with a unified mathematical form, which takes into account the multiple physical factors. The first term on the left-hand side represents the coupling force of the bubble oscillation, migration and ambient flow field; the second term is the source term due to the phase transition. The right-hand side represents the volume acceleration of the bubble. The enthalpy difference $H$ exhibits good extensibility, determined by the specific physical problems. Once $H$, $\dot m$ and $v$ are obtained, Eq. (\ref{unif}) can be solved.
As $\dot m = 0$, the above equation is simplified to 

\begin{equation}
\label{unif11}
\left( {\frac{{C - \dot R}}{R} + \frac{{\rm{d}}}{{{\rm{d}}t}}} \right)\left [{\frac{{{R^2}}}{C}\left( {\frac{1}{2}{{{\dot R}}^2} + \frac{1}{4}{{v}^2} + H} \right)} \right] = 2R\dot R_{}^2 + {R^2}\ddot R .
\end{equation}

The above equation simplifies to the Keller-Miksis equation when the bubble migration velocity is removed, and transforms to the Gilmore equation by making simple substitutions in the expansion. It can be simplified to the Rayleigh-Plesset equation if fluid compressibility is further neglected.

%
%
%
 
\subsection{Phase transition modelling}  
The key to solving the enthalpy difference $H$ at the bubble surface is to obtain the liquid pressure at the bubble surface, which is close to the phase transition. According to \citet{fa80}, the pressure balance on the surface of the bubble can be expressed as:  

\begin{equation}
\label{Pb}
{{P}_{\rm b}}={{P}_{\rm g}}-\frac{2\sigma }{R}-\frac{4\mu }{R}\left( \dot{R}-\frac{{\dot{m}}}{\rho } \right)-{{\dot{m}}^{2}}\left( \frac{1}{\rho }-\frac{1}{{{\rho }_{\rm g}}} \right),
\end{equation}
in which ${{P}_{\rm g}}$ is inner gas pressure; $\sigma$ is surface tension coefficient and $\mu$ is viscosity coefficient; ${\rho }_{\rm g}$ is average gas density.
 
	
 The bubble contents consist mainly of non-condensable gases and vapor \citep{bhl02}, which are considered to be uniformly distributed inside the bubble. 
 Considering that the gases inside the bubble are violently compressed at the end of the collapse, the van der Waals equation \citep{Yasui2021a,RN1317} is employed to model the uniform inner pressure of the bubble: 
\begin{equation}
\label{Pg}
\left( {{P}_{\rm g}}+\frac{a}{{{\upsilon_{\rm m} }^{2}}} \right)\left( \upsilon_{\rm m} -b \right)={{R}_{\rm g}}T,
\end{equation}
where $\upsilon_{\rm m}$ is the molar volume $\upsilon_{\rm m} = N_{\rm A}V/n_{\rm t}$ ($N_{\rm A}$ is Avogadro number; $V$ is the volume of the bubble; $n_{\rm t}$ denotes the number of molecules inside the bubble); $T$ is the temperature at the bubble center; $R_{\rm g}$ is the gas constant; $a$ and $b$ are van der Waals constants with the following expressions:

\begin{equation}
\label{a_cal}
\left\{ {\begin{array}{*{20}{c}}
	{a = {{\left( {\sqrt {{a_{\rm a}}} \frac{{{n_{\rm a}}}}{{{n_{\rm t}}}} + \sqrt {{a_{\rm v}}} \frac{{{n_{\rm v}}}}{{{n_{\rm t}}}}} \right)}^2}}\\
	{b = {{\left( {\sqrt {{b_{\rm a}}} \frac{{{n_{\rm a}}}}{{{n_{\rm t}}}} + \sqrt {{b_{\rm v}}} \frac{{{n_{\rm v}}}}{{{n_{\rm t}}}}} \right)}^2}}
	\end{array}} \right.,
\end{equation}
where ${n}_{\rm a}$ and ${n}_{\rm v}$ are the numbers of air and vapor molecules, respectively; ${n_{\rm t}}={n}_{\rm a}+{n}_{\rm v}$; ${a}_{\rm a}$ and ${a}_{\rm v}$ are the van der Waals force of air and vapor molecules, respectively; ${b}_{\rm a}$ and ${b}_{\rm v}$ are the volumes occupied by air molecules and vapor molecules, respectively. As the van der Waals constants ($a$ and $b$) equal to zero, Eq. (\ref{Pg}) simplifies to the ideal gas equation. The change rate in the number of vapor molecules can be calculated as
\begin{equation}
\label{dn}
{{{\dot{n}}}_{\rm v}}=\frac{4\pi {{R}^{2}}\dot{m}{{N}_{\rm A}}}{{{M}_{\rm {mv}}}},
\end{equation}
where $M_{\rm {mv}}$ is molar mass of vapor.

A modified Hertz–Knudsen–Langmuir relationship \citep{Schrage1953,Akhatov2001} could be used to compute the net evaporation rate of mass:
\begin{equation}
\label{dm}
\dot{m}=\frac{{{\alpha }_{\rm m}}}{\sqrt{2\pi {{R}_{\rm v}}}}\left( \frac{P_{\rm s}}{\sqrt{{{T}_{\rm l}}(R)}}-\frac{\Gamma {{P}_{\rm v}}}{\sqrt{{{T}_{\rm B}}}} \right),
\end{equation}
where ${\alpha }_{\rm m}$ is the adaptation factor, and its value is characterized by the evaporation and condensation (the value of ${\alpha }_{\rm m}$ is taken to around 0.04 according to the previous works on the acoustic bubbles \citep{Yasui1998}, but it is not a fixed value for bubbles generated by different methods; it is determined depending on the specific properties of the bubble contents).
 $R_{\rm v}$ is the gas constant for vapor. $T_{\rm l}(R)$ denotes the liquid temperature at the bubble surface and $T_{\rm B}$ is the gas temperature at the bubble surface. $P_{\rm s}$ is the saturated vapor pressure at the temperature $T_{\rm l}(R)$. 
 $P_{\rm v}$ is the actual saturated vapor pressure. Following \citet{Rui2023}, $\Gamma$ is a correction factor taken as 1.0. Additionally, for bubbles where phase transition effects are minimal but there is a flow of gases into and out of the bubble, such as in the case of air-gun bubbles \citep{RN1335,shuai20}, the value of $\dot m$ in Eq. (\ref{dm}) can be computed in alternative forms according to different physical problems.


To describe the temperature change at the gas-liquid interface, two thermodynamic boundary layers are introduced according to the previous works \citep{fa80,Yasui1999}. Inside the bubble, the gas temperature varies from the temperature $T$ inside the bubble to $T_{\rm B}$ at the bubble surface. Outside the bubble, the liquid temperature $T_{\rm l}(r)$ changes from the temperature $T_{\rm l}(R)$ at the bubble surface to $T_{\infty}$ at infinity. Here, following the linear model proposed by \citet{Yasui2016}, the gas temperature distribution near the inner bubble surface could be described as follows:  

\begin{equation}
\label{pianT}
{{\left. \frac{\partial T}{\partial r} \right|}_{r=R}}=\frac{{{T}_{\rm B}}-T}{n\lambda },
\end{equation}
where $\lambda$ is the mean molecular free range of the gas and $n$ is a constant that determines the thickness of the thermodynamic boundary layer.

Assuming that there is a temperature jump at the gas-liquid interface from $T_{\rm B}$ to $T_{\rm l}(R)$ \citep{RN1341,RN1315}, the gas temperature $T_{\rm B}$ at the bubble surface is calculated by
\begin{equation}
\label{TB}
{{T}_{\rm B}}={{T}_{\rm l}(R)}-\frac{\kappa }{2kn'}\sqrt{\frac{\pi \bar{m}}{2k{{T}_{\rm B}}}}\frac{2-a'{{\alpha }_{\rm e}}}{{{\alpha }_{\rm e}}}{{\left. \frac{\partial T}{\partial r} \right|}_{r=R}},
\end{equation}
where $k$ denotes Boltzmann constant; $n'$ denotes the number density of molecules inside the bubble; ${\alpha }_{\rm e}$ is the thermodynamic coefficient of adaptation; $a'=0.827$; $\kappa$ is the thermal conductivity coefficient of water vapor, and $\bar{m}$ is the average mass of the molecules inside the bubble $\bar{m}={\left( {{n}_{\rm v}}{{M}_{\rm v}}+{{n}_{\rm a}}{{M}_{\rm a}} \right)}/{\left( {{n}_{\rm t}}{{N}_{\rm A}} \right)}\;$ (${M}_{\rm a}$ and ${M}_{\rm v}$ are the mass of air and vapor inside the bubble, respectively).



The temperature distributions within the thermodynamic boundary layer outside the bubble surface are extensively examined \citep{fa80, Tian2022,RN1370}, as they need to satisfy the thermodynamic boundary conditions both at the bubble surface and at infinity. Here, the exponential distribution model proposed by \citet{RN1313} is employed to describe the temperature gradient outside the bubble at the collapse stage:

\begin{equation}
\label{distribT}
{T_{\rm l}}\left( r \right) =\left\{ {\begin{array}{*{20}{c}}
	{ {T_\infty } + \left[ {{T_{\rm l}}\left( R \right) - {T_\infty }} \right]{e^{ - \frac{{r - R}}{{{T_\infty } - {T_{\rm l}}\left( R \right)}}{{\left. {\frac{{\partial {T_{\rm l}}}}{{\partial r}}} \right|}_{r = R}}}}, \text{ }{\rm when}{\rm{ }}\left[ {{T_{\rm l}}\left( R \right) - {T_\infty }} \right]{{\left. {\frac{{\partial {T_{\rm l}}}}{{\partial r}}} \right|}_{r = R}} < 0{\rm{ }}}\\
	{ {T_\infty } + A{e^{ - B{{\left( {r - Y} \right)}^2}}},\text{ } {\rm{    when  }}\left[ {{T_{\rm l}}\left( R \right) - {T_\infty }} \right]{{\left. {\frac{{\partial {T_{\rm l}}}}{{\partial r}}} \right|}_{r = R}} > 0}
	\end{array}} \right.,
\end{equation}
where $A$, $B$ and $Y$ are computed as 

\begin{equation}
\label{para}
A = \left[ {{T_{\rm l}}\left( R \right) - {T_\infty }} \right]{e^{Be_1^2}},B = \frac{{{{\left. {\frac{{\partial {T_{\rm l}}}}{{\partial r}}} \right|}_{r = R}}}}{{2{e_1}\left[ {{T_{\rm l}}\left( R \right) - {T_\infty }} \right]}},Y = R + {e_1},{e_1} = e'\left| {\frac{{{T_{\rm l}}\left( R \right) - {T_\infty }}}{{{{\left. {\frac{{\partial {T_{\rm l}}}}{{\partial r}}} \right|}_{r = R}}}}} \right|,
\end{equation}
where $e'=0.001$.

Then, according to the energy conservation within the thermodynamic boundary layer outside the bubble, the variation of $T_l(R)$ with time could be updated as

\begin{equation}
\label{TII}
\frac{4}{3}\pi \rho {c_{\rm p}}\left[ {{{\left( {R + {\delta _{\rm e}}} \right)}^3} - {R^3}} \right] \frac{\partial {T_{\rm l}}(R)}{\partial t} = 4\pi {R^2}\left( { - {\kappa_{\rm l}}{{\left. {\frac{{\partial {T_{\rm l}}}}{{\partial r}}} \right|}_{r = R}}} \right) - 4\pi {\left( {R + {\delta _{\rm e}}} \right)^2}\left( { - {\kappa_{\rm l}}{{\left. {\frac{{\partial {T_{\rm l}}}}{{\partial r}}} \right|}_{r = R + {\delta _{\rm e}}}}} \right) ,
\end{equation}
where $c_{\rm p}$ denotes the specific heat of liquids at constant pressure; $\kappa_{\rm l}$ is the thermal conductivity of liquids; $\delta_{\rm e}$ is the thickness \citep{RN1313,RN1317} of the thermodynamic boundary layer outside the bubble:

\begin{equation}
\label{deltaII}
{\delta _{\rm e}} =\left\{ {\begin{array}{*{20}{c}}
	{ {{\left[ {{T_{\rm l}}\left( R \right) - {T_\infty }} \right]} \mathord{\left/
				{\vphantom {{\left[ {{T_{\rm l}}\left( R \right) - {T_\infty }} \right]} {{{\left. {\frac{{\partial {T_{\rm l}}}}{{\partial r}}} \right|}_{r = R}}}}} \right.
				\kern-\nulldelimiterspace} {{{\left. {\frac{{\partial {T_{\rm l}}}}{{\partial r}}} \right|}_{r = R}}}},\text{ }{\rm{   when  }}\left[ {{T_{\rm l}}\left( R \right) - {T_\infty }} \right]{{\left. {\frac{{\partial {T_{\rm l}}}}{{\partial r}}} \right|}_{r = R}} < 0{\rm{ }}}\\
	{  {e_1} + 1/\sqrt B ,\text{ }{\rm{    when  }}\left[ {{T_{\rm l}}\left( R \right) - {T_\infty }} \right]{{\left. {\frac{{\partial {T_{\rm l}}}}{{\partial r}}} \right|}_{r = R}} > 0}
	\end{array}} \right.,
\end{equation}
and ${\left. {{{\partial {T_{\rm l}}} \mathord{\left/
				{\vphantom {{\partial {T_{\rm l}}} {\partial r}}} \right.
				\kern-\nulldelimiterspace} {\partial r}}} \right|_{r = R + {\delta _{\rm e}}}}$ is calculated using Eq. (\ref{distribT}) and (\ref{para}); the temperature gradient of liquid at the bubble surface ${\left. {{{\partial {T_{\rm l}}} \mathord{\left/
				{\vphantom {{\partial {T_{\rm l}}} {\partial r}}} \right.
				\kern-\nulldelimiterspace} {\partial r}}} \right|_{r = R}}$ is determined by the continuity condition \citep{fa80} of the heat flux:

\begin{equation}
\label{energyflux}
{\kappa_{\rm l}}{\left. {\frac{{\partial {T_{\rm l}}}}{{\partial r}}} \right|_{r = R}} = {\kappa}{\left. {\frac{{\partial {T}}}{{\partial r}}} \right|_{r = R}} + {\dot m}L,
\end{equation}
where $L$ is latent heat.

The temperature change inside the bubble with respective to time can be updated according to the change of internal energy
\begin{equation}
\label{dTT}
M {c}_{\rm v} {\dot{T}} = \dot E,
\end{equation}
where 
$M$ is the gas mass inside the bubble; ${c}_{\rm v}$ is the average specific heat capacity of the gas inside the bubble. Following \citet{Zhong2020} and \citet{RN1339}, the change rate of energy is computed as 

\begin{equation}
\label{dE}
\begin{aligned}
\dot{E}=&-S_{\rm b}\dot{R}{{P}_{\rm g}}+\frac{S_{\rm b}\left[ {{{\dot{m}}}_{\text{e}}}{{e}_{\rm v}}\left( {{T}_{\rm l}} \right)-{{{\dot{m}}}_{\text{c}}}{{e}_{\rm v}}\left( {{T}_{\rm B}} \right) \right]{{N}_{\rm A}}}{{{M}_{\rm {mv}}}}  +S_{\rm b}\kappa {{\left. \frac{\partial T}{\partial r} \right|}_{r=R}} \\ 
& \text{    } \text{    }+S_{\rm b}{{\sigma }_{\rm r}}\left( T_{\rm B}^{4}-{{T}^{4}} \right)  \\ 
\end{aligned},
\end{equation}
in which ${e}_{\rm v}$ is the energy carried by a vapor molecule; ${\sigma }_{\rm r}$ is the Stefan-Boltzmann constant. The first term on the right-hand side is the work done by the bubble on the surrounding fluids, the second term represents the energy carried by the evaporation and condensation of the fluids, the third term is the energy produced by heat conduction, and the fourth term is the energy produced by heat radiation. In some studies \citep{RN1335,Nagalingam2023,Jiacheng2024}, the two thermodynamic boundary layers at the gas-liquid interface are often ignored for simplicity. The gas temperature at the inner surface of the bubble is replaced with the temperature at the bubble center, while the liquid  temperature at the outer surface of the bubble is substituted with the ambient temperature. Then, the change rate of energy can be expressed simply as 

\begin{equation}
\label{dE11}
\dot{E}= -S_{\rm b}\dot{R}{{P}_{\rm g}}+{S_{\rm b}}\left({{c}_{\rm p}}-{{c}_{\rm v}}\right){{T}}\dot{m}-S_{\rm b}\kappa_{\rm s} \left( {{T}}-{{T}_{\infty}} \right),
\end{equation}
where $\kappa_{\rm s}$ is a heat transfer coefficient.

Here, some values of the parameters at room temperature \citep{Zhong2020} involved in the above model are provided as shown in Table \ref{param}. 

\begin{table}	
	\centering		
\caption{Values of partial thermodynamic parameters at room temperature (293 K) }	 
		\begin{tabular}{c|c|c} 
		\toprule
			\textbf{Name} & \textbf{Variant} & \textbf{Value} \\
	Surface tension coefficient& $\sigma$& $0.075$  $\ {\rm N/m} $\\	
Viscosity coefficient& $\mu$& $0.001$  $\ {\rm Pa \cdot s}$\\				
			Avogadro constant& $N_{\rm A}$& 6.02$\times 10^{23}\ $ ${\rm {mol}}^{-1}$ \\
			Gas constant& $R_{\rm g}$& 8.314 $\ {\rm {J}} \cdot {\rm{mol^{-1}}} \cdot {\rm{K^{-1}}}$ \\
			Gas constant of vapor& $R_{\rm v}$& 461 $\ {\rm {J}} \cdot {\rm{Kg^{-1}}} \cdot {\rm{K^{-1}}}$ \\
			van der Waals force of air molecules& $a_{\rm a}$& 0.1402 $\ {\rm {J}} \cdot {\rm{m^{3}}} \cdot {\rm{mol^{-2}}}$ \\
			van der Waals force of vapor molecules& $a_{\rm v}$& 0.5536 $\ {\rm {J}} \cdot {\rm{m^{3}}} \cdot {\rm{mol^{-2}}}$ \\
			Volume occupied by air molecules& $b_{\rm a}$& $3.753 \times 10^{-5}$ $\ {\rm{m^{3}}} \cdot {\rm{mol^{-1}}}$ \\
			Volume occupied by vapor molecules& $b_{\rm v}$& $3.049 \times 10^{-5}$ $\ {\rm{m^{3}}} \cdot {\rm{mol^{-1}}}$ \\
			Thermal accommodation coefficient& $\alpha_{\rm e}$& $1$   \\
			Thermal conductivity of vapor& $\kappa$& $0.02$  $\ W \cdot \rm{m^{-1}} \cdot {\rm{K^{-1}}}$  \\
			Thermal conductivity of water& $\kappa_{\rm l}$& $0.55$  $\ W \cdot \rm{m^{-1}} \cdot {\rm{K^{-1}}}$  \\
			Boltzmann constant& $k$& $1.38 \times 10^{-23}$  $\ \rm{J} \cdot \rm{K^{-1}} $  \\
			Stefan-Boltzmann constant& $\sigma_{\rm r}$& $5.67 \times 10^{-8}$  $\ \rm{W} \cdot \rm{m^{-2}} \cdot \rm{K^{-4}} $  \\
			Latent heat of water& $L$& $2.4 \times 10^{6}$  $\ \rm{J} \cdot \rm{Kg^{-1}} $  \\	
		\end{tabular}
		\label{param}
\end{table}

\subsection{Bubble migration equation} 

In this section, the bubble migration equation is derived to solve the migration velocity in Eq. (\ref{unif}). As the bubble migrates along the axis of $\theta=0$, the kinetic boundary condition for the bubble migration can be expressed as 
\begin{equation}\label{eq-kinetic-condition-mig}
 \frac{\mathrm{d}}{\mathrm{d}t}\int_V r\cos\theta \mathrm{d}V = \frac{4}{3}\pi {R^3}{ v}.
\end{equation}

According to Reynold's transport theorem, the above equation can be expanded as

\begin{equation}\label{eq-kinetic-condition-mig12}
 \int_V {\frac{{\partial \left( {r\cos \theta } \right)}}{{\partial t}}{\rm{d}}V}  +  \int_S {u_{r}R\cos \theta {\rm{d}}S}  = \frac{4}{3}\pi {R^3}{ v}.
\end{equation}

The first term at the left side in Eq. (\ref{eq-kinetic-condition-mig12}) disappears because the integrant does not change over time. Combining the above equation with Eq. (\ref{eq-phi-r}) yields
\begin{equation}\label{eq-kinetic-condition-mig1}
q+\frac{R}{C}q' =\frac{1}{2}{ {v} }{{R}^{3}}.
\end{equation}

Let $Q(t) = q(t - R/C)$, then we have
\begin{equation}
\label{QQ}
\frac{\rm{d}\textit{Q}}{\rm{d}\textit{t}}=\left( {1 - \frac{{\dot R}}{C}} \right)q'{|_{r = R}}.
\end{equation}

Combining the above two equations and differentiating Eq. (\ref{eq-kinetic-condition-mig1}) with respect to time gives
\begin{equation}
\label{Q}
\left( \frac{C-\dot{R}}{R}+\frac{\rm d}{{\rm d}t} \right)\left( \frac{R}{C-\dot{R}}\frac{{\rm d}Q}{{\rm d}t} \right) =\frac{1}{2} {\dot{  v}} {{R}^{3}}+\frac{3}{2}{{R}^{2}}\dot{R}{  {v} }.
\end{equation}

Eq. (\ref{Q}) is the bubble migration equation in a unified mathematical form. The left-hand side of Eq. (\ref{Q}) represents the migration force exerted on the bubble by the flow field, while the right-hand side is the change rate of the bubble's momentum with respect to time. Similar to the bubble oscillation equation (\ref{eq15}), once we have determined ${{\rm d}Q}/{{\rm d}t}$, we can obtain the migration equation for the bubble.
To determine ${{\rm d}Q}/{{\rm d}t}$, 
we first set out the momentum equation for the bubble:
\begin{equation}
\label{forceeq}
\frac{{\rm d}\left( M{\bm v}_{\rm m} \right)}{{\rm d}t}= M{\bm g} -\int_{S}{{{P}_{\rm b}}{\bm n}\text{d}S}-\frac{1}{2}\pi {{R}^{2}}\rho{{C}_{\rm d}}\mathbb{C}\left( \bm{v} \right),
\end{equation}
where the three terms on the right-hand side denote the gravity, the inertial force, and the drag force of the bubble, respectively; $C_{\rm d}$ is the drag coefficient; $\mathbb{C}({\bm x})={\bm x}|\bm x|$.

Multiplying the Bernoulli equation (\ref{eq-bernoulli}) by $\bm n$ and integrating it on the bubble surface with $H$ retaining the zero-order term $(P_{\rm b}-P_{\rm a})/\rho$, then the terms containing $\theta$ vanish. Consequently, the inertial force can be obtained as
\begin{equation}
\label{inertiaforce}
\begin{array}{l}
\int_S {{P_{\rm b}}{\bm n}{\rm{d}}S}  = \int_S {\left[ {{P_{\rm{a}}} - \rho {{\left. {\left( {\frac{{\partial \varphi }}{{\partial t}} + \left( {v\cos \theta , - v\sin \theta } \right) \cdot \left( {{u_{\rm r}},{u_{\rm {\theta}} }} \right) + \frac{1}{2}\left( {u_{\rm r}^2 + u_{\rm \theta} ^2} \right)} \right)} \right|}_{r = R}}} \right]{\bm n}{\rm{d}}S} \\
= \int_V {\nabla {P_{\rm{a}}}{\rm d}V}  - \rho \int\limits_0^\pi  {\left( \begin{array}{l}
	{\left. {\frac{{\partial \varphi }}{{\partial t}}} \right|_{r = R}} + \left( {v\cos \theta , - v\sin \theta } \right) \cdot \left( {{u_{\rm r}},{u_{\rm \theta} }} \right)\\
	+ \frac{1}{2}\left( {u_{\rm r}^2 + u_{\rm \theta} ^2} \right)
	\end{array} \right){\bm e} \cdot 2\pi {R^2}\cos \theta \sin \theta {\rm d}\theta } \\
= \frac{4}{3}\pi \rho \left( {q' + \frac{R}{C}q''} \right){\bm e} + \frac{4}{3} \pi R^3 \nabla {P_{\rm{a}}}
\end{array},
\end{equation}
where $\nabla {P_\mathrm{a}}=\rho {\bm g}$ in the free field with gravity.

Substituting Eq. (\ref{inertiaforce}) into Eq. (\ref{forceeq}) and organizing Eq. (\ref{forceeq}) gives
\begin{equation}
\label{forceeq1}
\left(q' + \frac{R}{C}q''\right) {\bm e}=  - \frac{{{R^3}\nabla {P_{\text{a}}}}}{\rho } - \frac{3}{8}{R^2}{C_{\rm d}}\mathbb{C}\left( {\bm v} \right) + \frac{{{\rho _{\rm g}}}}{\rho }{R^3}{\bm g} - \frac{{{\rho _{\rm g}}}}{\rho }{R^3}{\dot {\bm v}}_{\rm m} - \frac{{3{R^2}\dot m}}{\rho }{\bm v}_{\rm m}.
\end{equation}

 Multiply both sides of Eq. (\ref{eq-kinetic-condition-mig1}) by $\bm e$ and derive it with respect to time. Then, associating it with Eq. (\ref{forceeq1}), we can eliminate $q''$ to have 
\begin{equation}
\label{forceeq12}
\frac{{\dot R}}{{C - \dot R}}q'{\bm e} = \frac{C}{{C - \dot R}}\left( {\frac{1}{2}\dot {\bm v}{R^3} + \frac{3}{2}{\bm v}{R^2}\dot R} \right) - \left( { - \frac{{{R^3}\nabla {P_{\text{a}}}}}{\rho } - \frac{3}{8}{R^2}{C_{\rm d}}\mathbb{C}\left( \bm v \right) + \frac{{{\rho _g}}}{\rho }{R^3}{\bm g} - \frac{{{\rho _g}}}{\rho }{R^3}{\dot {\bm v}}_{\rm m} - \frac{{3{R^2}\dot m}}{\rho }{\bm v}_{\rm m}} \right).
\end{equation}

By associating Eq. (\ref{QQ}) with the above equation, the expression of ${{\rm d}Q}/{{\rm d}t}$ is obtained

\begin{equation}
\label{forceeq13}
\begin{aligned}
\frac{\text{d}Q}{\text{d}t}{\bm e}&=\frac{{C - \dot R}}{{\dot R}}\left( {\frac{1}{2}\dot {\bm v}{R^3} + \frac{3}{2}{\bm v}{R^2}\dot R} \right) \\ &+ \frac{{{{\left( {C - \dot R} \right)}^2}}}{{C\dot R}}\left( {  \frac{{{R^3}\nabla {P_{\text{a}}}}}{\rho } + \frac{3}{8}{R^2}{C_{\rm d}}\mathbb{C}\left( \bm v \right) - \frac{{{\rho _{\rm g}}}}{\rho }{R^3}{\bm g} + \frac{{{\rho _{\rm g}}}}{\rho }{R^3}{\dot {\bm v}}_{\rm m} + \frac{{3{R^2}\dot m}}{\rho }{\bm v}_{\rm m}} \right).
\end{aligned}
\end{equation}	

Similarly, multiplying both sides of Eq. (\ref{Q}) by $\bm e$ and associating it with Eq. (\ref{forceeq13}), we can arrive at the bubble migration equation: 
\begin{equation}
\label{migration}
\begin{aligned}
& \left[ 1-\frac{R\ddot{R}}{\left( C-\dot{R} \right)\dot{R}}+\frac{R}{C-\dot{R}}\frac{\text{d}}{\text{d}t} \right]\left( \frac{1}{2}{{R}^{3}}\dot{\bm v}+\frac{3}{2}{{R}^{2}}\dot{R}{\bm v} \right) \\ 
& =\left[ 1-\frac{R\ddot{R}}{\left( C-\dot{R} \right)\dot{R}}+\frac{R}{C}\frac{\text{d}}{\text{d}t} \right]\left[\frac{{{\rho }_{\rm g}}}{\rho }R^3\left({\bm g} - {\dot{\bm v}}_{\rm m} \right) -3{{R}^{2}}\frac{{\dot{m}}}{\rho }{\bm v}_{\rm m}  -\frac{{{R}^{3}}\nabla {{P}_{\rm a}}}{\rho }-\frac{3}{8}{{C}_{\text{d}}}{{R}^{2}}\mathbb{C}\left( \bm v \right)\right].  
\end{aligned}
\end{equation}

In the free field, ${\bm u}_{\rm a}=0$ and thus ${\bm v}_{\rm m}={\bm v}$. When we consider the non-spherical bubble oscillation in many cases, the added mass coefficient of the bubble $C_{\rm a}$ needs to be introduced in Eq. (\ref{migration}). In fact, in the above equations, the added mass coefficient of the bubble is implicitly fixed to 0.5 due to the assumption of spherical bubbles. By analogy with the derivation of the added mass force of the bubble in an incompressible flow, the expression in the small parentheses on the left-hand side of Eq. (\ref{migration}) can be rewritten as $(C_{\rm a}R^3\dot{{\bm v}} + 3C_{\rm a}R^2\dot{R} {\bm v})$ if $C_{\rm a}$ is not equal to 0.5. 
%
When the high-order terms related to fluid compressibility are neglected, the above equation simplifies to

\begin{equation}
\label{migration111}
 C_{\rm a}{{R}}\dot{\bm v}+3 C_{\rm a} \dot{R}{\bm v} +\frac{{{R}}\nabla {{P}_{\rm a}}}{\rho }+\frac{3}{8}{{C}_{\text{d}}}\mathbb{C}\left( \bm v \right) +\frac{3{\dot{m}}}{\rho }{\bm v}_{\rm m} -\frac{{{\rho }_{\rm g}}}{\rho }R\left({\bm g} - {\dot{\bm v}}_{\rm m} \right) 
 = {\bm 0}.  
\end{equation}

Eq. (\ref{migration111}) could be simplified to the form in our previous works \citep{unified2023} if the last two terms on the left-hand side representing the phase transition and inertia effect of the internal gas are ignored.

\subsection{Multiple-bubble interaction and boundary effects}

In this section, we incorporate the effects of multiple bubbles and boundaries into the present theoretical model.
 The principle involves modifying the ambient pressure $P_{\rm a}$ and velocity ${\bm u}_{\rm a}$ of the background flow field of the bubble when the effect of multiple bubbles is considered. Also, accounting for boundary effects is achieved by introducing image bubbles, thereby transforming it into a multiple-bubble problem. Firstly, we provide the pressure and velocity in the flow field induced by a single bubble. Differentiating Eq. (\ref{eq-velocity-potential}) with respect to $\boldsymbol{r}$ and $t$ with the velocity potential of bubble migration ignored, and combining it with Eq. (\ref{dfdt}), we can establish the correlation between the physical information at $|\bm r|$ and the bubble surface:

	\begin{equation}
	\boldsymbol{u}{(\boldsymbol{r},t)} = 
	-\frac{\boldsymbol{o} - \boldsymbol{r}}{|\boldsymbol{o} - \boldsymbol{r}|^3}\left.
	\left[R^2 \left(\dot{R}-\frac{\dot m}{\rho} \right)-\frac{R}{C}\left(|\boldsymbol{o} - \boldsymbol{r}|-R\right)\left(H+\frac12\left(\dot{R}-\frac{\dot m}{\rho}\right)^2+\frac14{v}^2\right)	
\right]
	\right|_{\left({R},t_{\rm c}\right)},
	\end{equation}

\noindent and

\begin{equation}
\frac{\partial \varphi{(\boldsymbol{r},t)}}{\partial t} = 
-\left. \frac{R}{|\boldsymbol{o} - \boldsymbol{r}|} \left( H + \frac12 \left(\dot{R}-\frac{\dot m}{\rho}\right)^2+\frac14{v}^2\right) \right|_{\left({R},t_{\rm c}\right)},
\end{equation}

\noindent where $ t_{\rm c} = t - (|{\boldsymbol{r}}| - R)/C $, and it denotes the initiation moment of a disturbance induced by the bubble surface that later arrives at $ \boldsymbol{r} $ at $ t $. The flow pressure induced by the bubble can be solved by substituting the above two equations into the Bernoulli equation:

 \begin{equation}
p =  - \rho \frac{{\partial \varphi \left( {{\bm r},t} \right)}}{{\partial t}} - \frac{1}{2} \rho |{\bm u}\left( {{\bm r},t} \right)|^2 + {{{P_\infty }}}.
 \end{equation}

%
%
%
%
%

Assuming there are $U$ bubbles in the flow field, the velocity and pressure of the background field for bubble $N$ can be expressed as:

\begin{equation}
\label{eq447}
\boldsymbol{u}_{\mathrm{a}}{\left( {{{\boldsymbol{o}}_{\rm N}},t} \right)}=
\sum_{\genfrac{}{}{0pt}{2}{G=1,U}{G\ne N}} \boldsymbol{u}_{\rm G}{(\boldsymbol{o}_{\rm N},t)},
\end{equation}

\noindent and

\begin{equation}
	\label{eq-pa-boundary4}
	P_{\mathrm{a}}{(\boldsymbol{o}_{\rm N},t)} =
	- 
	\rho \sum_{\genfrac{}{}{0pt}{2}{G=1,U}{G\ne N}} \frac{\partial  {\varphi}_{\rm G}{(\boldsymbol{o}_{\rm N},t)}}{\partial t} 
	-\frac12 \rho |\boldsymbol{u}_{\mathrm{a}}{(\boldsymbol{o}_{\rm N},t)}|^2 + P_{\infty},
\end{equation}
respectively. The dynamics of multiple bubbles can be addressed by incorporating the above two equations into the oscillation equation and migration equation.

Further, assuming that a infinite flat boundary exists near the bubble, defined by ${\bm r} \cdot {\boldsymbol{e}}_{\rm{b}}+s=0$ (where ${\boldsymbol{e}}_{\rm{b}}$ is the outward unit normal vector of the boundary plane and $s$ is a constant), the position of the image bubble $N_{\rm i}$ of bubble $N$ about the boundary could satisfy ${{\boldsymbol{o}}_{\rm N_i}} = {{\boldsymbol{o}}_{\rm N}} - 2 \left( {{{\boldsymbol{o}}_{\rm N}} \cdot {{\boldsymbol{e}}_{\rm{b}}}  + s} \right){{\boldsymbol{e}}_{\rm{b}}}$. The size and oscillation velocity of bubble $N$ and $N_{\rm i}$ always remain exactly the same, while the position and migration velocity of the two bubbles are always symmetric about the boundary plane.
A reflection coefficient $\xi $ is used to determine the property of the boundary. Specifically, $\xi = 1.0$ for a rigid boundary, and $\xi = -1$ for an ideal free surface. 
Therefore, when the bubble $N$ is affected by other bubbles and the boundary, the velocity and pressure of the flow field at $\boldsymbol{o}_{\rm N}$ can be expressed as

\begin{equation}
\label{eq47}
\boldsymbol{u}_{\mathrm{a}}{\left( {{{\boldsymbol{o}}_{\rm N}},t} \right)}=
\sum_{\genfrac{}{}{0pt}{2}{G=1,U}{G\ne N}} \boldsymbol{u}_{\rm G}{(\boldsymbol{o}_{\rm N},t)} + \xi \sum_{G=1,U} \boldsymbol{u}_{{\rm G_i}}{(\boldsymbol{o}_{\rm N},t)},
\end{equation}

\noindent and

\begin{small}
	\begin{equation}
	\label{eq-pa-boundary}
	P_{\mathrm{a}}{(\boldsymbol{o}_{\rm N},t)} =
	- 
	\rho \sum_{\genfrac{}{}{0pt}{2}{G=1,U}{G\ne N}} \frac{\partial  {\varphi}_{\rm G}{(\boldsymbol{o}_{\rm N},t)}}{\partial t} 
	-\rho \xi \sum_{G=1,U} \frac{\partial {\varphi}_{{\rm G_i}}{(\boldsymbol{o}_{\rm N},t)}}{\partial t}
	-\frac12 \rho |\boldsymbol{u}_{\mathrm{a}}{(\boldsymbol{o}_{\rm N},t)}|^2 + P_{\infty},
	\end{equation}
\end{small}
respectively.

The dynamics of bubble $N$ near the boundary and other bubbles can be addressed by incorporating Eq. (\ref{eq47}) and Eq. (\ref{eq-pa-boundary}) into the oscillation equation and migration equation of bubble $N$. 

\begin{figure}
	\centering\includegraphics[width=1.0\linewidth]{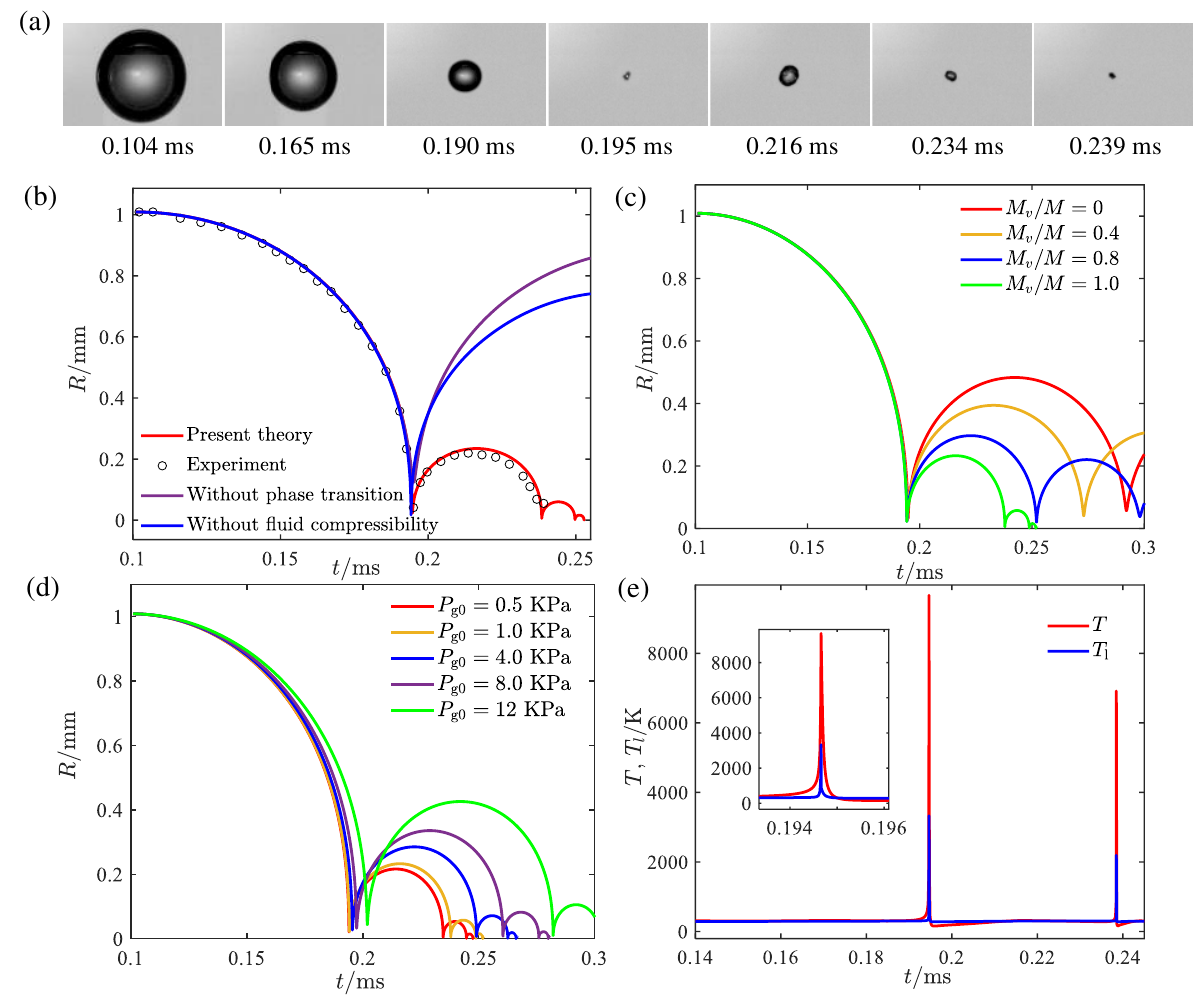}
	\caption{Laser bubble experiment in the free field and its comparison with theoretical results. (a) High-speed photography images of the bubble oscillation over time. Frame width: 3.45 mm. (b) Comparison of the bubble radius between theory and experiment ($R_0 = 1.01$ mm, $P_{\rm g0} = 1.0$ KPa, $M_{\rm v}/M=1.0 $, $\alpha_{\rm m} = 0.064$). (c) Effect of initial proportion of vapor inside the bubble. (d) Effect of initial internal bubble pressure. (e) Temporal variations of the temperature at the bubble center $T$ and the liquid temperature at the bubble surface $T_{\rm l}$.}
	\label{laserfree}
\end{figure}

\section{Validation of the present theoretical model}
\label{3}


In this section, we conduct experiments on bubbles with different sources and environmental conditions, capturing the bubble oscillation and migration processes. The experimental results are compared with the theoretical values to validate the present theoretical model.


\subsection{Bubble dynamics in the free field}	
	
Firstly, we validate the present theoretical model through two cavitation bubble experiments in a free field. In the first experiment, the bubble is generated by laser focusing with the maximum bubble radius of 1.01 mm, and the experimental setup can be referred to in the previous work \citep{RN1269}. Fig. \ref{laserfree}(a) shows high-speed photography images in the first two oscillation cycles of the bubble, where the bubble remains nearly spherical during the first cycle and undergoes slight deformation in the second cycle. Fig. \ref{laserfree}(b) presents a comparison between the computed bubble radius and the experimental data. The theoretical calculations start from the moment the bubble reaches its maximum radius, at which point the oscillation velocity of the bubble is zero. Since the initial conditions of bubbles in the experiments are difficult to determine, we discuss the effects of initial parameters here. The temperature of the fluid domain $T_{\infty}$ is fixed at 293 K in all the cases unless otherwise stated. The initial vapor proportion $M_{\rm v}/M$ and the internal pressure $P_{\rm g0}$ of the bubble are two important parameters that significantly affect the maximum radius of the bubble during the second cycle. Fig. \ref{laserfree}(c) shows the effect of the initial vapor proportion on the bubble radius at a fixed $P_{\rm g0}$, and Fig. \ref{laserfree}(d) shows the effect of the initial internal pressure at a fixed initial $M_{\rm v}/M$. A higher initial vapor proportion and a lower initial internal pressure significantly reduce the maximum radius of the bubble during the second cycle. Considering the small content of non-condensable gases inside the laser bubble \citep{Liang2022}, the initial vapor proportion is set at 1.0 in this case. The initial internal pressure of the bubble is set at 1.0 KPa to match the experimental bubble radius in the second cycle, and $\alpha_{\rm m}$ in the Hertz–Knudsen–Langmuir relationship is chosen as 0.064. The number of air and vapor molecules at the initial moment is estimated by the ideal gas equation. The present theoretical model effectively captures the experimental bubble radius in the first two cycles. In addition, Fig. \ref{laserfree}(b) also shows the calculation results without the phase transition and without the fluid compressibility, respectively. The results indicate that both phase transition and fluid compressibility are important factors affecting the energy loss of bubbles, with phase transition having a more significant impact on laser-induced bubbles. Fig. \ref{laserfree}(e) illustrates the temporal variation of the temperature of the bubble center and the liquid temperature at the bubble surface. During the majority of the bubble cycle, the temperature inside the bubble and at the bubble surface remain in close proximity. 
At the final stages of the bubble collapse, the temperature at the bubble center rises more rapidly over time compared to that at the bubble surface.
The discrepancy between the two temperatures primarily manifests during the intense phase of bubble collapse, where there is a stark rise from the ambient water temperature as the bubble collapses, followed by a precipitous drop toward equilibrium. 

\begin{figure}
	\centering\includegraphics[width=1.0\linewidth]{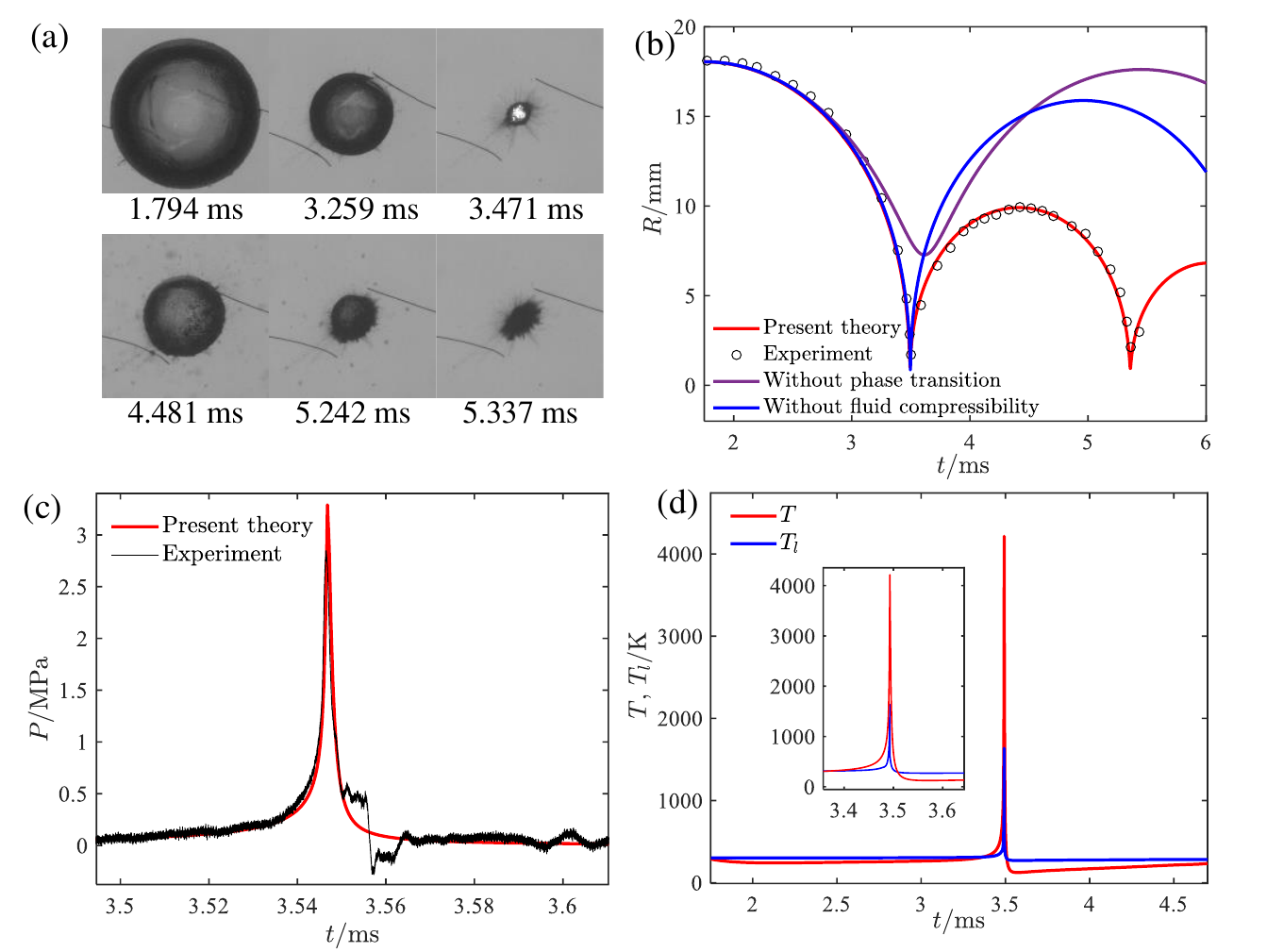}
	\caption{Spark-generated bubble experiment in the free field and its comparison with theoretical results. (a) High-speed photography images of the bubble oscillation over time. Frame width: 4.43 cm. (b) Comparison of the bubble radius between theory and experiment ($R_0=18.1$ mm, $P_{\rm g0} = 12$ KPa, $M_{\rm v}/M=1.0 $, $\alpha_{\rm m} = 0.043$). (c) Comparison of the flow-field pressure induced by bubble oscillation between theory and experiment. (d) Temporal variations of the temperature at the bubble center $T$ and the liquid temperature at the bubble surface $T_{\rm l}$.}
	\label{sparkfree}
\end{figure}

In the second experiment, the bubble is generated by underwater electrical discharge, reaching a maximum bubble radius of 18.1mm. The experiment method can be referred to in the work of \citet{Hanrui2022}. Fig. \ref{sparkfree}(a) shows the temporal evolution of the bubble shape. The bubble is accompanied by a large amount of flocculent impurities at the end of the second cycle, making it difficult to clearly observe the bubble profile. However, in general, the profile of the bubble at the moment of maximum volume is clear enough to accurately obtain the maximum radius of the bubble during the first two cycles, allowing for calculations using the present theoretical model. Fig. \ref{sparkfree}(b) and \ref{sparkfree}(c) compare the bubble radius and the flow-field pressure induced by the bubble oscillation, respectively. The flow-field pressure is measured by a PCB free-field sensor placed 8.5 cm away from the bubble center. Similar to Fig. \ref{laserfree}, the initial vapor proportion of the bubble is set at 1.0. The initial internal pressure of the bubble is 12 KPa, and $\alpha_{\rm m}=0.043$. The smaller peak of the flow-field pressure in the experiment may be attributed to the limited sampling frequency of the sensor or slight changes in the sensor's position under the influence of bubble oscillation. Overall, the theoretical calculations well reproduce the bubble radius and the oscillation pressure in the experiment. Neglecting the effect of phase transition and fluid compressibility significantly overestimates the radius of the bubble during the second cycle. Fig. \ref{sparkfree}(d) shows the temporal variation of the gas temperature at the center of the bubble and the liquid temperature at the surface of the bubble. Compared to the laser-induced bubble, the internal gas of the spark-generated bubble reaches a lower temperature at the moment of minimum volume, which to some extent indicates that the collapse intensity of the bubble is weaker, resulting in less energy loss of the bubble at the end of the first cycle. Note that the maximum bubble displacement of the bubble center in the first two cycles of the cases in this section is small enough to ignore the bubble migration. Thus, the migration features of the bubbles are not discussed here.

\begin{figure}
	\centering\includegraphics[width=1.0\linewidth]{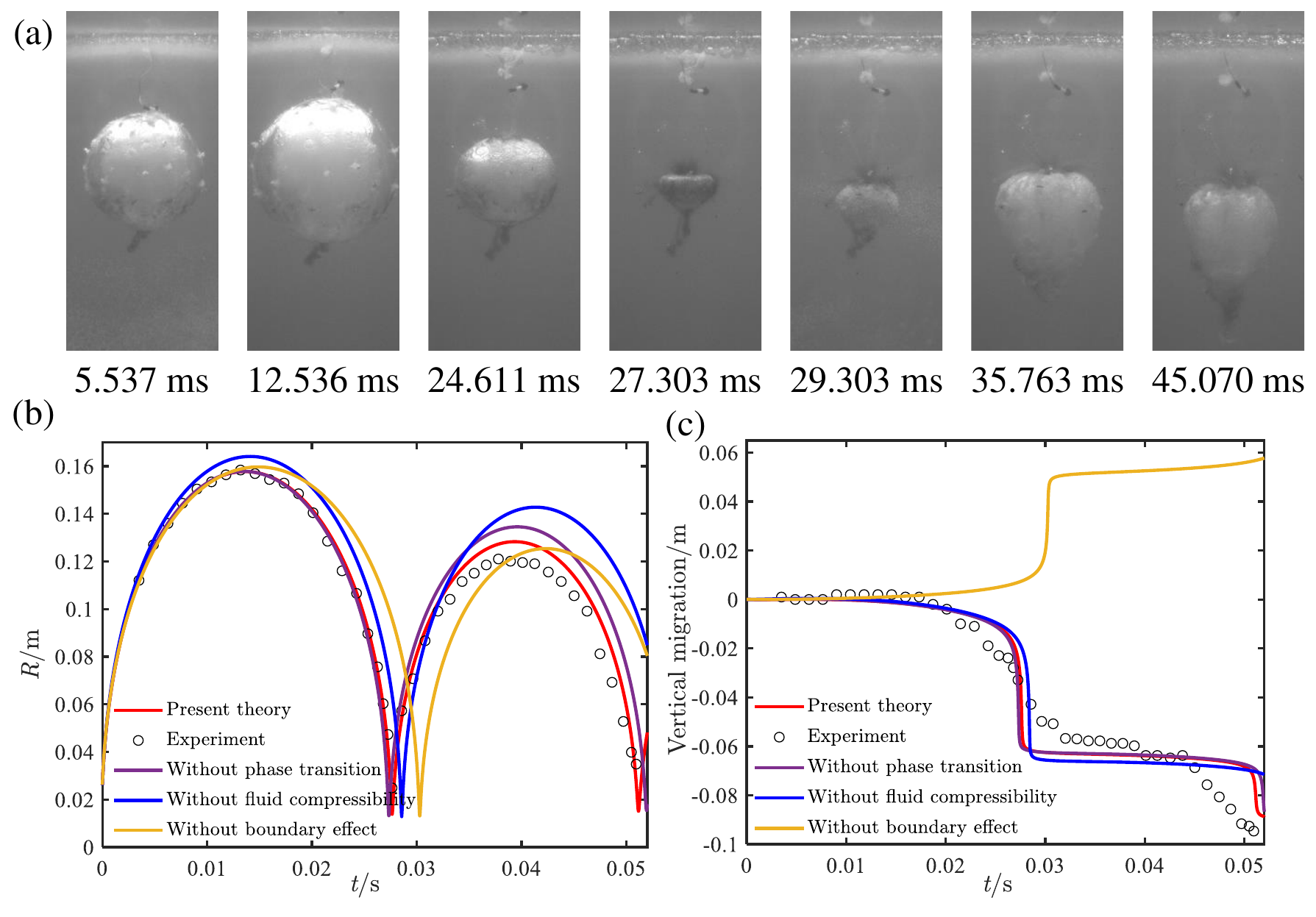}
	\caption{{ Experiment of an underwater explosion bubble near the free surface and its comparison with theoretical results.} {({a})} High-speed photography images of the bubble oscillation over time. Frame width: 0.510 m. {({b})} Comparison of the bubble radius between the theoretical and the experimental results (${R_{0}}=0.026$ m, ${\dot{R}_{0}}=109$ m/s, ${P_{\rm g0}}=2.74$ MPa, $M_{\rm v}/M=0.01 $, $\alpha_{\rm m}=0.041$). {({c})} Comparisons of the bubble migration between the theoretical and the experimental results.}
	\label{Contour3}
\end{figure}

\subsection{Bubble dynamics under different boundary conditions}

In this section, we validate the effects of boundaries and multiple bubbles in the bubble equation through two bubble experiments. The first case is an underwater explosion bubble generated by 1.05 g TNT explosives near the free surface. The experiment is conducted in a cubic water tank, which can be referred to in the work of \citet{unified2023}.
 The underwater explosion bubble is initially $30$ cm from the free surface with a maximum radius of $15.8$ cm. The initial condition of the bubble is calculated according to the shock wave theory in the previous works \citep{unified2023}: ${R_{0}}=0.026$ m, ${\dot{R}_{0}}=109$ m/s, ${P_{\rm g0}}=2.74$ MPa. The value of $\alpha_{\rm m}$ is 0.041. In this case, we set the initial vapor proportion in the bubble content as $1\%$ in theory. This can be explained by the presence of a large amount of non-condensable contents inside the underwater explosion bubble. Meanwhile, we provide the calculated results without the effect of phase transition, showing that the phase transition plays a relatively minimal role for underwater explosion bubbles. The effects of the boundary and liquid compressibility play an important role in bubble dynamics. Removing the boundary effect in theory leads to a significantly larger bubble oscillation period and the wrong migration direction because the free surface could accelerate the bubble oscillation and induce the bubble to migrate downwards. As the boundary effect is neglected, the bubble migration is controlled by the buoyancy of the bubble. Removing the liquid compressibility results in a great deviation from the experimental values for both the maximum bubble radius and the energy loss of the bubble. 

\begin{figure}
	\centering\includegraphics[width=1.0\linewidth]{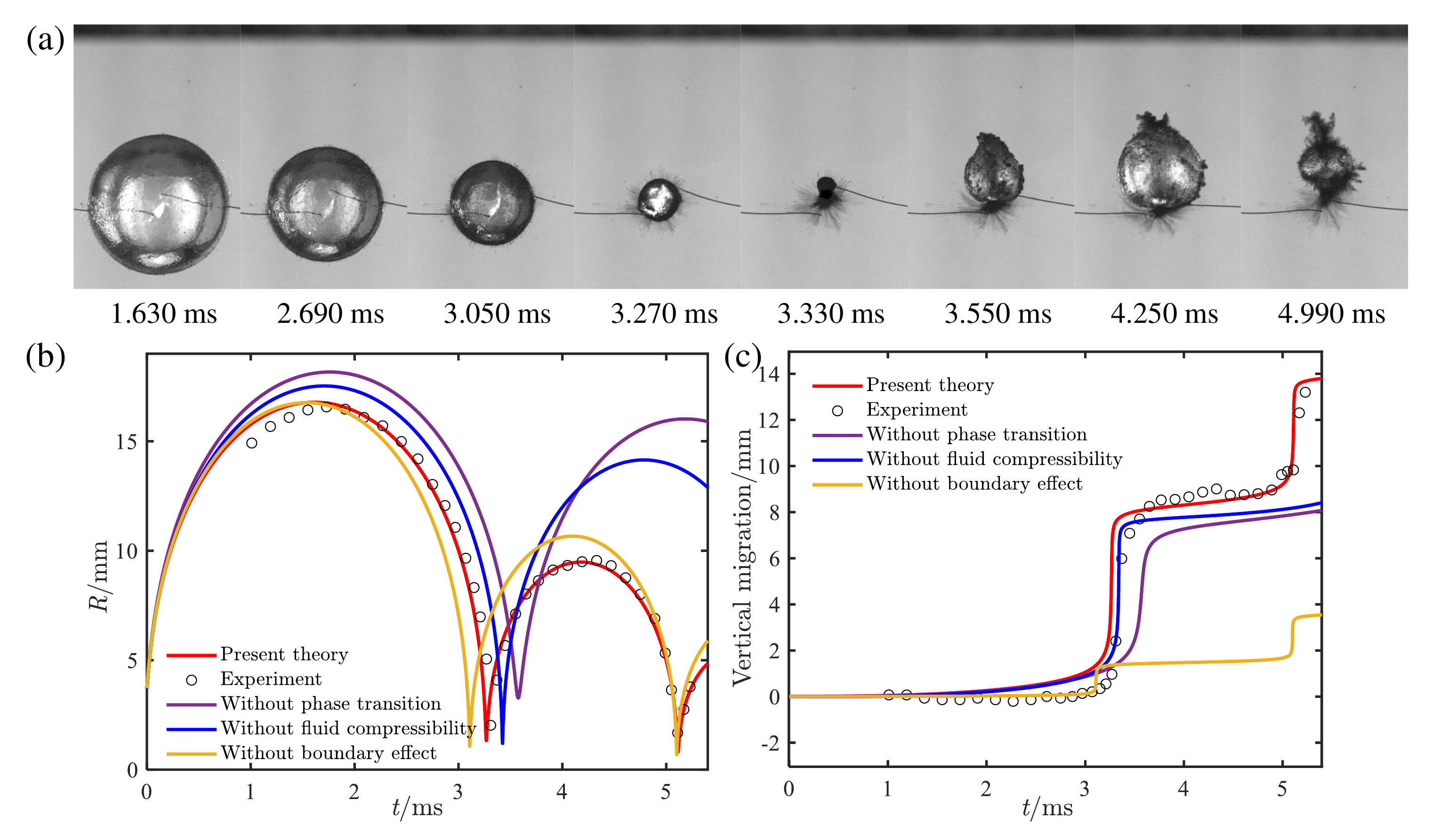}
	\caption{{Spark-generated bubble experiment near a rigid wall and its comparison with theoretical results.} {({a})} High-speed photography images of the bubble oscillation over time. Frame width: 40 mm. {({b})} Comparison of the bubble radius between the theoretical and the experimental results (${R_{0}}=3.69$ mm, ${\dot{R}_{0}}=60$ m/s, ${P_{\rm g0}}=5.0$ MPa, $M_{\rm v}/M=1.0 $, $\alpha_{\rm m}=0.043$). {({c})} Comparison of the bubble migration between the theoretical and the experimental results.  }
	\label{Contour1}
\end{figure}

The second case is a spark-generated bubble near the above wall. Fig. \ref{Contour1} provides the temporal progression of the bubble shape, accompanied by a comparison between the theoretical and experimental results of the bubble radius and vertical displacements.
 The bubble is 44 mm from the above wall at inception, and the maximum bubble radius is 16.6 mm. The bubble does not have a tendency to migrate in most of the first cycle, but migrates upwards obviously at the end of collapse and the second cycle due to the effect of the wall. To compute the dynamics of a spark-generated bubble from its inception, the initial conditions of the bubble are obtained by integrating backward from the moment of maximum bubble volume \citep{w13,unified2023}. The detailed procedure is as follows: the computation begins at the moment of maximum bubble volume when the bubble radius is known and its oscillation velocity is zero; next, the internal pressure at the moment of maximum bubble volume depends on the experimental bubble radius in the second cycle; the calculation then proceeds in reverse along the time axis from the moment of maximum bubble volume until the computed time approaches zero.
According to the method, the initial conditions of the bubble are: ${R_{0}}=3.69$ mm, ${\dot{R}_{0}}=60$ m/s, ${P_{\rm g0}}=5.0$ MPa. The initial vapor proportion in theory and the value of $\alpha_{\rm m}$ are the same as the case of the spark-generated bubble in Fig. \ref{sparkfree}.
The impact of various physical factors is also analyzed in this case.
Removing the boundary effect causes the bubble period to decrease due to the presence of the wall. The amplitude of the bubble migration is significantly weaker when the wall is neglected. 
The energy loss of the bubble is much weaker when the phase transition is not considered compared to the computational results without the fluid compressibility. As indicated by the previous underwater explosion experiments and this case, the relative impact of phase transition and fluid compressibility on the energy loss of bubbles is closely related to the composition of gases.

\subsection{Multiple bubble dynamics}
\begin{figure}
	\centering\includegraphics[width=1.0\linewidth]{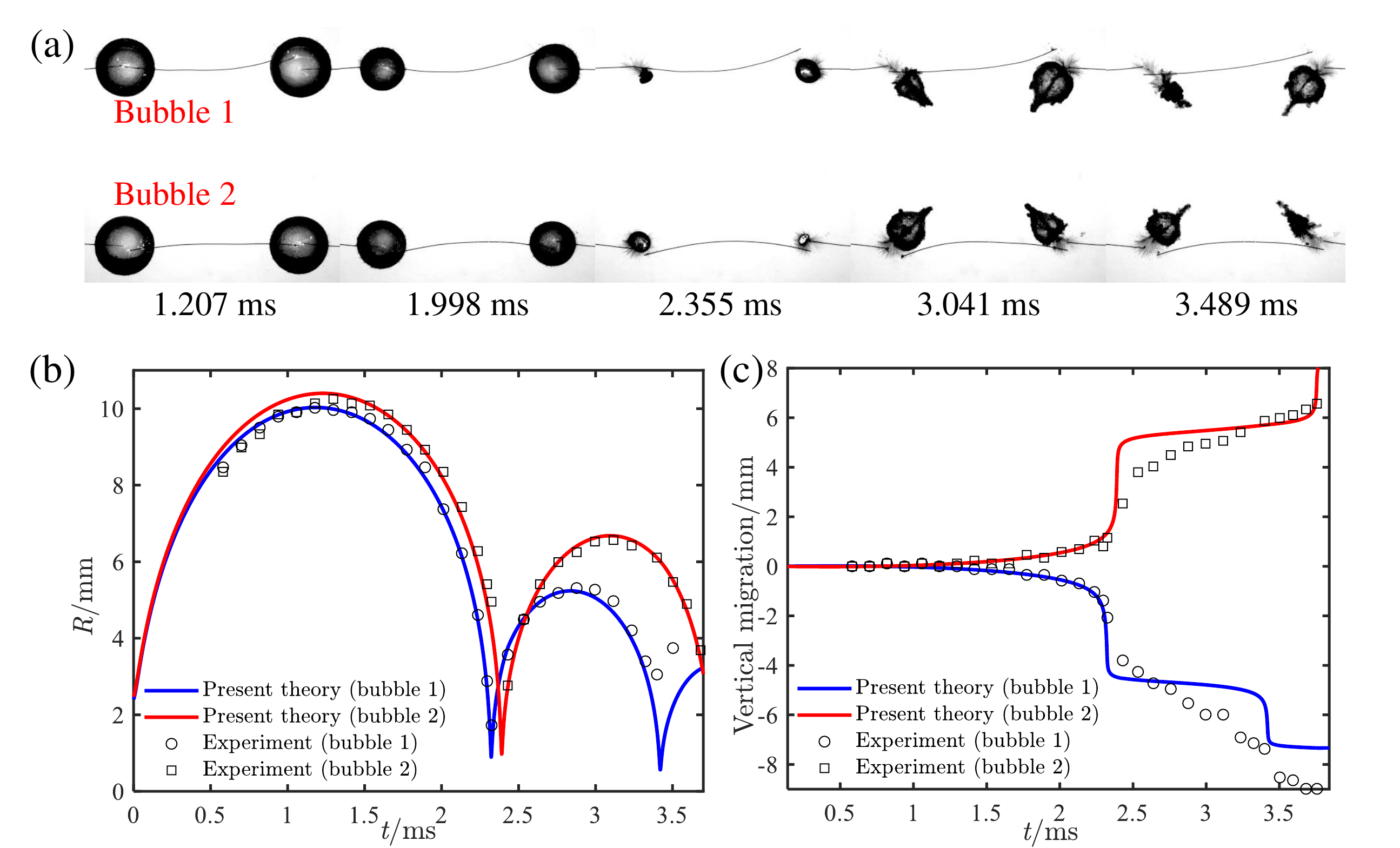}
			\caption{{ Comparison of the interaction of four spark-generated bubbles with the theoretical results. ({a}) High-speed photography images of the bubble oscillation over time. Frame width: 162 mm. ({b}) Comparison of the bubble radius between the theoretical and the experimental results ($R_{01}=1.04 \rm{}$ mm,
					$R_{02}=1.06 \rm{}$ mm, $\dot{R}_{01}=\dot{R}_{02}=10$ m/s, and $P_{\rm g01}=P_{\rm g02}=40 $ MPa, $M_{\rm v}/M=0.99 $, $\alpha_{\rm m}=0.041$). ({c}) Comparison  of the vertical bubble migration between the theoretical and the experimental results.}  }
	\label{Contour2}
\end{figure}

Finally, we carry out an experiment with multiple spark-generated bubbles and compared it with the theoretical results, as shown in Fig. \ref{Contour2}. Four bubbles are generated simultaneously at the four vertices of a square plane with a side length of 60 mm, as shown in Fig. \ref{Contour2}(a). The maximum radius of the upper-left and lower-right bubbles is 10.0 mm, and that of the remaining two bubbles is 10.3 mm. We denote the upper-left bubble as bubble 1, and the lower-left bubble as bubble 2. The initial oscillation conditions for the two bubbles are: $R_{01}=1.04 \rm{}$ mm,
$R_{02}=1.06 \rm{}$ mm, $\dot{R}_{01}=\dot{R}_{02}=10$ m/s, and $P_{\rm g01}=P_{\rm g02}=40 $ MPa. Bubble 1 shares the same initial conditions as the lower right bubble, while the two other bubbles also have identical initial conditions. This is considering that the corresponding two bubbles remain symmetrical for the majority of the bubble oscillation cycle.
 The proportion of vapor in the bubble contents is 0.99 and $\alpha_{\rm m}=0.041$. During the bubble oscillation, the four bubbles migrate towards the center of the square plane due to the mutual attraction among bubbles. Fig. \ref{Contour2}(b) and \ref{Contour2}(c) compare the radius and vertical displacement of the two bubbles, respectively. Overall, our theoretical model well reproduces the bubble radius and displacement in the experiment. 
 It is observed that the experimental bubble radius at the end of the second cycle is larger than the computed values. This discrepancy may be attributed to the measurement errors caused by the frame rate of the high-speed camera and the perturbation of frothy impurities on the bubble.

\section{Discussion on the energy loss of bubbles}
\label{4}

\begin{figure}
	\centering\includegraphics[width=1.0\linewidth]{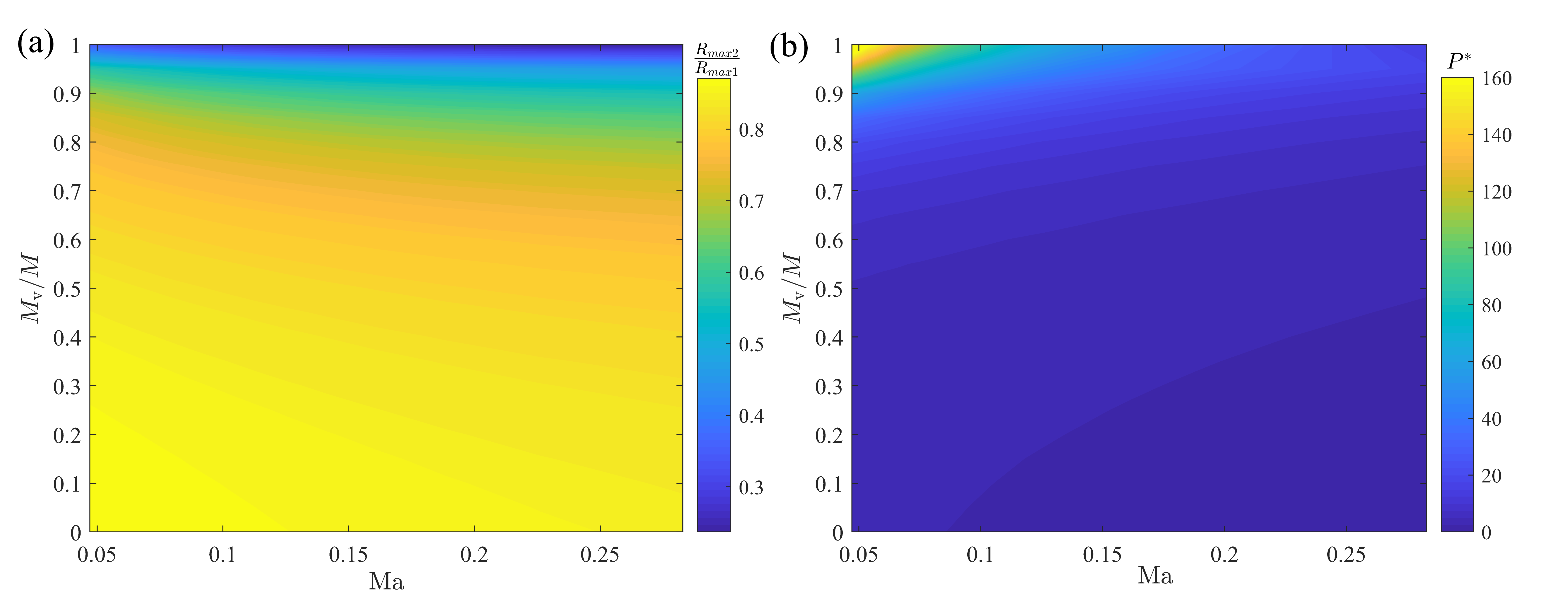}
	\caption{Distribution of (a) radius ratio during the first two cycles of the bubble and (b) scaled internal bubble pressure at the moment of minimum bubble volume for varying Mach numbers and vapor proportions (${{R}_{0}}^*=0.19$, ${\dot{R}_{0}}^*=0$, ${{P}_{\rm g0}}^*=50$).}
	\label{T-percent}
\end{figure}

In this section, we examine the influence of phase transition on the energy loss of bubbles. Firstly, we present the distribution of feature parameters of bubbles across different Mach numbers ${\rm Ma}=\sqrt{P_{\rm g0}/\rho}/C$ and initial vapor proportions $M_{\rm v}/M$, as depicted in Fig. \ref{T-percent}. The Mach number serves to quantify the impact of fluid compressibility, and the vapor proportion represents the influence of phase transition. In this section, all physical quantities, with the exception of the temperature, are rendered dimensionless by using the maximum radius of the bubble $R_{\rm max}$, the density of the liquid $\rho$, and the hydrostatic pressure at the bubble's initial location $P_{\infty}$.
The dimensionless physical quantities are indicated by the superscript `*' in the latter descriptions. The studied characteristic parameters are the radius ratio during the first two cycles $R_{\rm max2}/R_{\rm max1}$ and the scaled internal bubble pressure $P^*=P^*_{\rm max} r^*$ ($P^*_{\rm max}$ and $r^*$ are the peak pressure inside the bubble and the minimum bubble radius at the first collapse stage, respectively). $R_{\rm max2}/R_{\rm max1}$ is used to measure the energy loss of the bubble in the first two cycles. $P^*$ is utilized to characterize the energy at the moment of minimum bubble volume, and it equals the pressure peak induced by the bubble at the unit distance. The initial conditions of bubbles are: ${{R}_{0}}^*=0.19$, ${\dot{R}_{0}}^*=0$, ${{P}_{\rm g0}}^*=50$. The value of $\alpha_{\rm m}$ equals to 0.041. 
The initial vapor proportion inside the bubble is altered under the condition of a constant initial internal pressure.
	 In the calculations, the effect of bubble migration is removed in order to analyze the effect of vapor proportion and Mach number more accurately.
 Fig. \ref{T-percent}(a) illustrates the distribution of $R_{\rm max2}/R_{\rm max1}$ for varying $\rm Ma$ and proportions of vapor $M_{\rm v}/M$. The energy loss of bubbles shows a steady increase as the Mach number and vapor proportion increase. Note that when the vapor proportion approaches 1, the energy loss of bubbles is substantially greater compared to other cases. By examining the relationship between bubble energy and radius $E_1/E_2=(R_{\rm max1}/R_{\rm max2})^3$ ($E_1$ and $E_2$ denote the bubble energy in the first and second cycle, respectively), it is observed that the energy loss in vapor bubbles is typically over 80\%, which is at least twice the energy loss observed in the bubbles formed by non-condensable gases at the same Mach number. Fig. \ref{T-percent}(b) shows the distribution of $P^*$ for varying $\rm Ma$ and proportions of vapor $M_{\rm v}/M$. $P^*$ decreases as the Mach number increases because the fluid compressibility damps the pressure induced by the bubble oscillation. However, the increase in the vapor proportion results in a progressive increase in $P^*$, indicating that the vapor bubble can generate higher pressure peaks in the flow field compared to those composed purely of non-condensable gases.


To elucidate the underlying mechanism behind the variation of $P^*$ with $M_{\rm v}/M$ in Fig. \ref{T-percent}, we conduct an energy analysis for a special case where the bubble contents consist purely of vapor ($M_{\rm v}/M=1$). The formulas for calculating the internal energy $E_{\rm i}$, potential energy $E_{\rm p}$, kinetic energy $E_{\rm k}$ and radiated acoustic energy $E_{\rm a}$ of a bubble \citep{Qianxi2016, shuai20} are given below:

\begin{equation}
\label{Ei}
{E_{\rm i}} = 3\frac{{{n_{\rm t}}}}{{{N_{\rm A}}}}{R_{\rm g}}T,
\end{equation}

\begin{equation}
\label{Ep}
{E_{\rm p}} = {P_\infty }V,
\end{equation}

\begin{equation}
\label{Ek}
\begin{array}{l}
{E_{\rm k}} = -\frac{1}{2}\rho \int\limits_S {{\varphi _{\rm f}}\frac{{\partial {\varphi _{\rm f}}}}{{\partial n}}{\rm d}S}  = \frac{1}{2}\rho 4\pi {R^2}\frac{{f\left( {t - R/C} \right)}}{R}\left( {\dot R - \frac{{\dot m}}{\rho }} \right)\\
=   2\rho \pi {R^2}\left( {\dot R - \frac{{\dot m}}{\rho }} \right)\left[ {R\left( {\dot R - \frac{{\dot m}}{\rho }} \right) - \frac{1}{C}\left( {\ddot R - \frac{{\ddot m}}{\rho }} \right){R^2} - \frac{{2R}}{C}{{\left( {\dot R - \frac{{\dot m}}{\rho }} \right)}^2}} \right]
\end{array},
\end{equation}

\begin{equation}
\label{Ea}
{E_{\rm a}} = \frac{\rho }{{4\pi C}}\left[ { \int\limits_0^t {{{\ddot V}^2}\left( t \right){\rm d}t+\dot V\left( 0 \right)\ddot V\left( 0 \right) - \dot V\left( t \right)\ddot V\left( t \right)} } \right],
\end{equation}

where $\dot V$ and $\ddot V$ are the first and second-order derivatives of bubble volume with respect to time, respectively; $\ddot m$ is the second order derivative of $m$ with respect to time; $f(t-R/C)$ in Eq. (\ref{Ek}) is solved by conducting the perturbation method on Eq. (\ref{kine1}).

\begin{figure}
	\centering\includegraphics[width=1.0\linewidth]{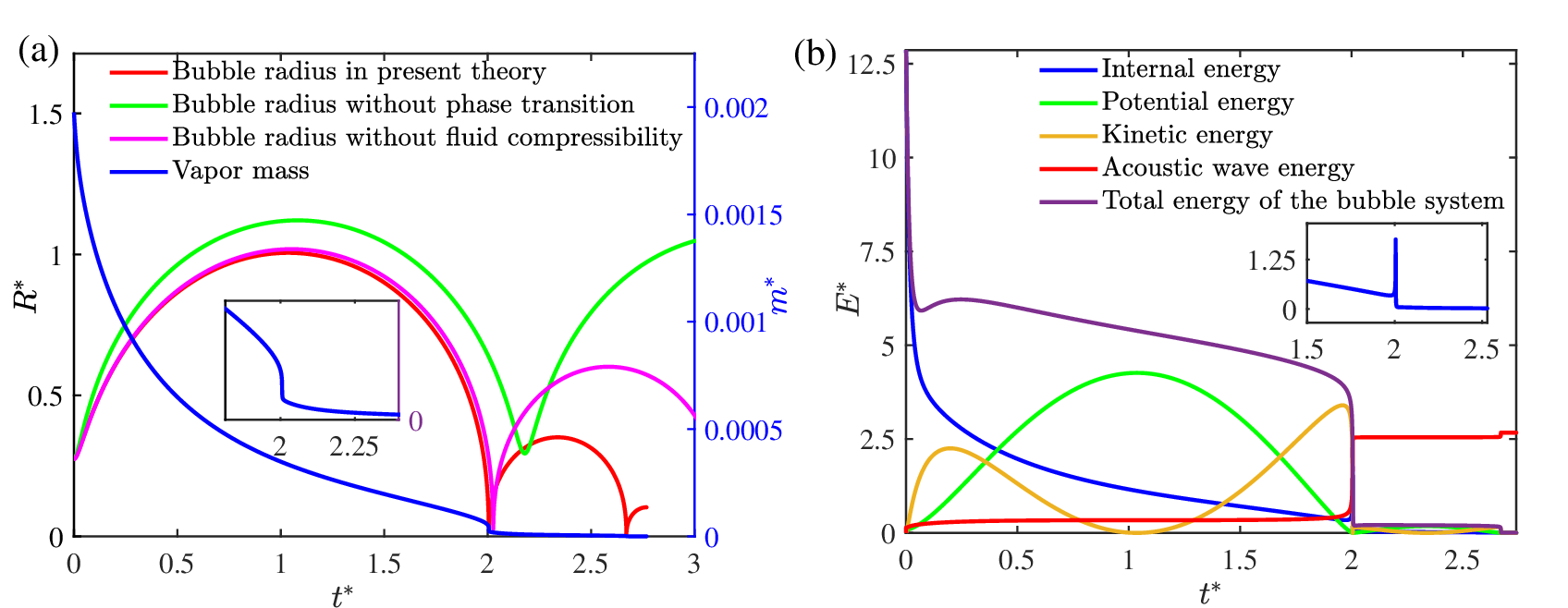}
	\caption{Time evolutions of mass and energy of a vapor bubble (${{R}_{0}}^*=0.19$, ${\dot{R}_{0}}^*=0$, ${{P}_{0}}^*=50$, $M_{\rm v}/M=1$, $\alpha_{\rm m}=0.041$). (a) Mass and radius of the bubble. (b) Various energies of the system.}
	\label{Energy}
\end{figure}

The time evolutions for the mass and radius of the bubble are presented in Fig. \ref{Energy}(a), along with the radius of the bubble in the absence of phase transition or fluid compressibility. The initial conditions of the bubble are: ${{R}_{0}}^*=0.19$, ${\dot{R}_{0}}^*=0$, ${{P}_{0}}^*=50$, $M_{\rm v}/M=1$, $\alpha_{\rm m}=0.041$. The condensation of vapor dominates the phase transition, leading to a consistent decreasing trend in the bubble mass. By comparing the results without considering the phase transition and without considering the compressibility of the fluid, it can be observed that the effect of the phase transition on the energy loss of the bubble is more pronounced. 
According to the relationship between bubble energy and radius, the energy loss of the bubble caused by the fluid compressibility ($E_2/E_1=0.87$) is less than $1/5$ that due to phase transition ($E_2/E_1=0.21$). 
Fig. \ref{Energy}(b) shows the time histories of the internal energy of the bubble, the kinetic energy and potential energy of fluids induced by bubble oscillation, and the acoustic wave energy that propagates away. The internal energy of the bubble decreases with time in most of the cycles, except for a small increase near the moment of minimum bubble volume due to the increase in temperature inside the bubble. The changes in the potential and kinetic energy of fluids are closely related to the volume and oscillation velocity of the bubble, but their amplitudes decrease significantly in the second bubble cycle compared to the first bubble cycle. The sum of the internal energy, the potential, and the kinetic energy characterizes the total energy of the bubble system, as shown by the purple solid line in Fig. \ref{Energy}(b). The total energy of the bubble system in the second cycle is reduced by about 4.7 near the moment of minimum bubble volume, while the acoustic wave energy radiated into the flow field during this period is about 2.4 (about 1/2 of the bubble energy loss). This contrasts with the trend suggested by the radius curves depicted in  Fig. \ref{Energy}(a). This feature implies that the impact of vapor condensation on bubble energy loss is manifested not only through the decrease in bubble mass but also through the intensification of the bubble collapse. The intensification results in a greater propagation of energy into the flow field in the form of acoustic radiation, which explains the observed increase of the scaled internal pressure with higher vapor proportions. 

\section{Conclusion}
\label{5}
In this study, the theoretical model for bubble dynamics that accounts for the oscillation, migration, phase transition, fluid compressibility, boundary effect, multiple bubbles,  viscosity, and surface tension is derived. The oscillation and migration equations are characterized by a unified mathematical form, and the terms in the equations have clear physical meanings.
The oscillation equation exhibits good extensibility and could be simplified to the classical Keller-Miksis equation after ignoring the effects of phase transition and bubble migration. 

The present theoretical model is validated through comparisons of the theoretical results with experimental values of a laser bubble and a spark-generated bubble in the free field. The effect of the initial vapor proportion and internal pressure on the bubble dynamics are discussed as important parameters for comparing theoretical results with experimental values. 
Subsequently, the theoretical model is further validated by an underwater explosion bubble experiment and several spark-generated bubble experiments under different boundary conditions. For underwater explosion bubbles, the non-condensable gases are the principal constituents of the bubble contents and the phase transition does not significantly affect the bubble dynamics; for laser bubbles and spark-generated bubbles, on the other hand, the phase transition is an important cause of bubble energy loss. 

Based on the present theoretical model, the effects of the Mach number and the initial vapor proportion inside the bubble on the energy loss of the bubble are investigated for an initially high-pressure bubble. The energy loss of the bubble increases with increasing Mach number and initial vapor proportion. 
Specifically, the energy loss of a vapor bubble is more than twice that of a bubble composed purely of non-condensable gases. Also, the radiated pressure peak by the bubble increases with the increasing vapor proportion. The vapor not only causes a loss of bubble contents through condensation but also leads to a more intense collapse. This intensified collapse, in turn, releases more acoustic energy into the surrounding fluid through pressure waves.

\bigskip
\noindent \textbf{Acknowledgments.} This work is funded by the National Natural Science Foundation of China (51925904, 52088102), the National Key R\&D Program of China (2022YFC2803500), Finance Science and Technology Project of Hainan Province (ZDKJ2021020). 

\bigskip
\noindent \textbf{Data availability statement.} The source code that support the findings of this study are openly available at {\color{blue}https://github.com/fslab-heu/new-bubble-theory}. 

\bigskip
\noindent \textbf{Declaration of Interests.} The authors report no conflict of interest.
\bibliography{sample}

\begin{thebibliography}{65}
\expandafter\ifx\csname natexlab\endcsname\relax\def\natexlab#1{#1}\fi
\providecommand{\bibinfo}[2]{#2}
\ifx\xfnm\relax \def\xfnm[#1]{\unskip,\space#1}\fi
\bibitem[{Klaseboer et~al.(2005)Klaseboer, Hung, Wang, Wang, Khoo, Boyce,
  Debono, and Charlier}]{khwwk}
\bibinfo{author}{E.~Klaseboer}, \bibinfo{author}{K.~C. Hung},
  \bibinfo{author}{C.~Wang}, \bibinfo{author}{C.~W. Wang},
  \bibinfo{author}{B.~C. Khoo}, \bibinfo{author}{P.~Boyce},
  \bibinfo{author}{S.~Debono}, \bibinfo{author}{H.~Charlier},
\newblock \bibinfo{title}{Experimental and numerical investigation of the
  dynamics of an underwater explosion bubble near a resilient/rigid structure},
\newblock \bibinfo{journal}{J. Fluid Mech.} \bibinfo{volume}{537}
  (\bibinfo{year}{2005}) \bibinfo{pages}{387--413}.
\bibitem[{Lyons et~al.(2019)Lyons, Haney, Fee, Wech, and Waythomas}]{lhfww19}
\bibinfo{author}{J.~Lyons}, \bibinfo{author}{M.~Haney},
  \bibinfo{author}{D.~Fee}, \bibinfo{author}{A.~Wech},
  \bibinfo{author}{C.~Waythomas},
\newblock \bibinfo{title}{Infrasound from giant bubbles during explosive
  submarine eruptions},
\newblock \bibinfo{journal}{Nat. Geosci.} \bibinfo{volume}{12}
  (\bibinfo{year}{2019}) \bibinfo{pages}{952--958}.
\bibitem[{Wang et~al.(2021)Wang, Gui, Zhang, Gao, Xu, and Jia}]{RN398}
\bibinfo{author}{S.~Wang}, \bibinfo{author}{Q.~Gui},
  \bibinfo{author}{J.~Zhang}, \bibinfo{author}{Y.~Gao},
  \bibinfo{author}{J.~Xu}, \bibinfo{author}{X.~Jia},
\newblock \bibinfo{title}{Theoretical and experimental study of bubble dynamics
  in underwater explosions},
\newblock \bibinfo{journal}{Phys. Fluids} \bibinfo{volume}{33}
  (\bibinfo{year}{2021}) \bibinfo{pages}{126113}.
\bibitem[{Versluis et~al.(2000)Versluis, Schmitz, von~der Heydt, and
  Lohse}]{vshl00}
\bibinfo{author}{M.~Versluis}, \bibinfo{author}{B.~Schmitz},
  \bibinfo{author}{A.~von~der Heydt}, \bibinfo{author}{D.~Lohse},
\newblock \bibinfo{title}{How snapping shrimp snap: Through cavitating
  bubbles},
\newblock \bibinfo{journal}{Science} \bibinfo{volume}{289}
  (\bibinfo{year}{2000}) \bibinfo{pages}{2114--2117}.
\bibitem[{Lohse et~al.(2001)Lohse, Schmitz, and Versluis}]{lsv01}
\bibinfo{author}{D.~Lohse}, \bibinfo{author}{B.~Schmitz},
  \bibinfo{author}{M.~Versluis},
\newblock \bibinfo{title}{Snapping shrimp make flashing bubbles},
\newblock \bibinfo{journal}{Nature} \bibinfo{volume}{413}
  (\bibinfo{year}{2001}) \bibinfo{pages}{477--478}.
\bibitem[{Lokhandwalla et~al.(2001)Lokhandwalla, McAteer, Jr, and
  Sturtevant}]{Lokhandwalla2001}
\bibinfo{author}{M.~Lokhandwalla}, \bibinfo{author}{J.~McAteer},
  \bibinfo{author}{J.~Jr}, \bibinfo{author}{B.~Sturtevant},
\newblock \bibinfo{title}{Mechanical haemolysis in shock wave lithotripsy
  (swl): Ii. in vitro cell lysis due to shear},
\newblock \bibinfo{journal}{Phys. Med. Biol.} \bibinfo{volume}{46}
  (\bibinfo{year}{2001}) \bibinfo{pages}{1245}.
\bibitem[{Ferrara et~al.(2007)Ferrara, Pollard, and Borden}]{fpb07}
\bibinfo{author}{K.~Ferrara}, \bibinfo{author}{R.~Pollard},
  \bibinfo{author}{M.~Borden},
\newblock \bibinfo{title}{Ultrasound microbubble contrast agents: Fundamentals
  and application to gene and drug delivery},
\newblock \bibinfo{journal}{Annu. Rev. Biomed.} \bibinfo{volume}{9}
  (\bibinfo{year}{2007}) \bibinfo{pages}{415--447}.
\bibitem[{Maeda and Colonius(2019)}]{Maeda2019}
\bibinfo{author}{K.~Maeda}, \bibinfo{author}{T.~Colonius},
\newblock \bibinfo{title}{Bubble cloud dynamics in an ultrasound field},
\newblock \bibinfo{journal}{J. Fluid Mech.} \bibinfo{volume}{862}
  (\bibinfo{year}{2019}) \bibinfo{pages}{1105--1134}.
\bibitem[{Verhaagen and Rivas(2016)}]{Verhaagen2016}
\bibinfo{author}{B.~Verhaagen}, \bibinfo{author}{D.~Rivas},
\newblock \bibinfo{title}{Measuring cavitation and its cleaning effect},
\newblock \bibinfo{journal}{Ultrason. Sonochem.} \bibinfo{volume}{29}
  (\bibinfo{year}{2016}) \bibinfo{pages}{619--28}.
\bibitem[{Oh et~al.(2018)Oh, Yoo, Seung, and Kwak}]{Jaekyoon2018}
\bibinfo{author}{J.~Oh}, \bibinfo{author}{Y.~Yoo}, \bibinfo{author}{S.~Seung},
  \bibinfo{author}{H.~Kwak},
\newblock \bibinfo{title}{Laser-induced bubble formation on a micro gold
  particle levitated in water under ultrasonic field},
\newblock \bibinfo{journal}{Exp. Therm. Fluid Sci.} \bibinfo{volume}{93}
  (\bibinfo{year}{2018}) \bibinfo{pages}{285--291}.
\bibitem[{Landel and Wilson(2021)}]{Landel2021}
\bibinfo{author}{J.~R. Landel}, \bibinfo{author}{D.~I. Wilson},
\newblock \bibinfo{title}{The fluid mechanics of cleaning and decontamination
  of surfaces},
\newblock \bibinfo{journal}{Annu. Rev. Fluid Mech.} \bibinfo{volume}{53}
  (\bibinfo{year}{2021}) \bibinfo{pages}{147--171}.
\bibitem[{Fujikawa and Akamatsu(1980)}]{fa80}
\bibinfo{author}{S.~Fujikawa}, \bibinfo{author}{T.~Akamatsu},
\newblock \bibinfo{title}{Effects of the non-equilibrium condensation of vapour
  on the pressure wave produced by the collapse of a bubble in a liquid},
\newblock \bibinfo{journal}{J. Fluid Mech.} \bibinfo{volume}{97}
  (\bibinfo{year}{1980}) \bibinfo{pages}{481--512}.
\bibitem[{Wang and Blake(2011)}]{Qianxi11}
\bibinfo{author}{Q.~Wang}, \bibinfo{author}{J.~R. Blake},
\newblock \bibinfo{title}{Non-spherical bubble dynamics in a compressible
  liquid. part 2. acoustic standing wave},
\newblock \bibinfo{journal}{J. Fluid Mech.} \bibinfo{volume}{679}
  (\bibinfo{year}{2011}) \bibinfo{pages}{559--581}.
\bibitem[{Brujan et~al.(2022)Brujan, Zhang, Liu, Ogasawara, and
  Takahira}]{Brujan2022}
\bibinfo{author}{E.~A. Brujan}, \bibinfo{author}{A.-M. Zhang},
  \bibinfo{author}{Y.-L. Liu}, \bibinfo{author}{T.~Ogasawara},
  \bibinfo{author}{H.~Takahira},
\newblock \bibinfo{title}{Jetting and migration of a laser-induced cavitation
  bubble in a rectangular channel},
\newblock \bibinfo{journal}{J. Fluid Mech.} \bibinfo{volume}{948}
  (\bibinfo{year}{2022}) \bibinfo{pages}{A6}.
\bibitem[{Preso et~al.(2024)Preso, Fuster, Sieber, Obreschkow, and
  Farhat}]{RN1349}
\bibinfo{author}{D.~B. Preso}, \bibinfo{author}{D.~Fuster},
  \bibinfo{author}{A.~B. Sieber}, \bibinfo{author}{D.~Obreschkow},
  \bibinfo{author}{M.~Farhat},
\newblock \bibinfo{title}{Vapor compression and energy dissipation in a
  collapsing laser-induced bubble},
\newblock \bibinfo{journal}{Phys. Fluids} \bibinfo{volume}{36}
  (\bibinfo{year}{2024}) \bibinfo{pages}{033342}.
\bibitem[{Rayleigh(1917)}]{r17}
\bibinfo{author}{L.~Rayleigh},
\newblock \bibinfo{title}{On the pressure developed in a liquid during the
  collapse of a spherical cavity},
\newblock \bibinfo{journal}{Philos. Mag.} \bibinfo{volume}{34}
  (\bibinfo{year}{1917}) \bibinfo{pages}{94--98}.
\bibitem[{Plesset(1949)}]{plesset49}
\bibinfo{author}{M.~Plesset},
\newblock \bibinfo{title}{The dynamics of cavitation bubbles},
\newblock \bibinfo{journal}{J. Appl. Mech.} \bibinfo{volume}{16}
  (\bibinfo{year}{1949}) \bibinfo{pages}{277--282}.
\bibitem[{Hicks(1970)}]{h70}
\bibinfo{author}{A.~N. Hicks}, \bibinfo{title}{Effect of bubble migration on
  explosion-induced whipping in ships}, \bibinfo{type}{Technical Report}
  \bibinfo{number}{3301}, Naval Ship Research and Development Center, Bethesda,
  MD, \bibinfo{year}{1970}.
\bibitem[{Best(1991)}]{RN272}
\bibinfo{author}{J.~P. Best}, \bibinfo{title}{The dynamics of underwater
  explosions}, \bibinfo{type}{Thesis}, \bibinfo{year}{1991}.
\bibitem[{Storey and Szeri(2001)}]{Storey2001}
\bibinfo{author}{B.~D. Storey}, \bibinfo{author}{A.~J. Szeri},
\newblock \bibinfo{title}{A reduced model of cavitation physics for use in
  sonochemistry},
\newblock \bibinfo{journal}{Proc. Math. Phys. Eng. Sci.} \bibinfo{volume}{457}
  (\bibinfo{year}{2001}) \bibinfo{pages}{1685--1700}.
\bibitem[{Oratis et~al.(2024)Oratis, Dijs, Lajoinie, Versluis, and
  Snoeijer}]{RN1356}
\bibinfo{author}{A.~T. Oratis}, \bibinfo{author}{K.~Dijs},
  \bibinfo{author}{G.~Lajoinie}, \bibinfo{author}{M.~Versluis},
  \bibinfo{author}{J.~H. Snoeijer},
\newblock \bibinfo{title}{A unifying rayleigh-plesset-type equation for bubbles
  in viscoelastic media},
\newblock \bibinfo{journal}{J. Acoust. Soc. Am.} \bibinfo{volume}{155}
  (\bibinfo{year}{2024}) \bibinfo{pages}{1593--1605}.
\bibitem[{Harkin et~al.(2001)Harkin, Kaper, and Nadim}]{Harkin2001}
\bibinfo{author}{A.~Harkin}, \bibinfo{author}{T.~Kaper},
  \bibinfo{author}{Nadim},
\newblock \bibinfo{title}{Coupled pulsation and translation of two gas bubbles
  in a liquid},
\newblock \bibinfo{journal}{J. Fluid Mech.} \bibinfo{volume}{445}
  (\bibinfo{year}{2001}) \bibinfo{pages}{377--411}.
\bibitem[{Bremond et~al.(2006)Bremond, Arora, Ohl, and Lohse}]{baol06}
\bibinfo{author}{N.~Bremond}, \bibinfo{author}{M.~Arora},
  \bibinfo{author}{C.-D. Ohl}, \bibinfo{author}{D.~Lohse},
\newblock \bibinfo{title}{Controlled multibubble surface cavitation},
\newblock \bibinfo{journal}{Phys. Rev. Lett.} \bibinfo{volume}{96}
  (\bibinfo{year}{2006}) \bibinfo{pages}{224501}.
\bibitem[{Herring(1941)}]{Herring1941}
\bibinfo{author}{C.~Herring}, \bibinfo{title}{Theory of the pulsations of the
  gas bubble produced by an underwater explosion}, \bibinfo{address}{Tech Rep.
  NDRC Division 6 Report C4-sr20. National Defense Research Committee},
  \bibinfo{year}{1941}.
\bibitem[{Gilmore(1952)}]{g52}
\bibinfo{author}{F.~R. Gilmore}, \bibinfo{title}{The growth and collapse of a
  spherical bubble in a viscous compressible liquid}, \bibinfo{type}{Report}
  \bibinfo{number}{26-4}, \bibinfo{year}{1952}.
\bibitem[{Keller and Kolodner(1956)}]{kk56}
\bibinfo{author}{J.~Keller}, \bibinfo{author}{I.~I. Kolodner},
\newblock \bibinfo{title}{Damping of underwater explosion bubble oscillations},
\newblock \bibinfo{journal}{J. Appl. Phys.} \bibinfo{volume}{27}
  (\bibinfo{year}{1956}) \bibinfo{pages}{1152}.
\bibitem[{Prosperetti and Lezzi(1986)}]{pp86}
\bibinfo{author}{A.~Prosperetti}, \bibinfo{author}{A.~Lezzi},
\newblock \bibinfo{title}{Bubble dynamics in a compressible liquid. {Part} 1.
  {First-order} theory},
\newblock \bibinfo{journal}{J. Fluid Mech.} \bibinfo{volume}{168}
  (\bibinfo{year}{1986}) \bibinfo{pages}{457--478}.
\bibitem[{Keller and Miksis(1980)}]{km80}
\bibinfo{author}{J.~Keller}, \bibinfo{author}{M.~Miksis},
\newblock \bibinfo{title}{Bubble oscillations of large amplitude},
\newblock \bibinfo{journal}{J. Acoust. Soc. Am.} \bibinfo{volume}{68}
  (\bibinfo{year}{1980}) \bibinfo{pages}{628}.
\bibitem[{Ma et~al.(2018)Ma, Hsiao, and Chahine}]{Chahine2018}
\bibinfo{author}{J.~Ma}, \bibinfo{author}{C.~T. Hsiao}, \bibinfo{author}{G.~L.
  Chahine},
\newblock \bibinfo{title}{Numerical study of acoustically driven bubble cloud
  dynamics near a rigid wall},
\newblock \bibinfo{journal}{Ultrason. Sonochem.} \bibinfo{volume}{40}
  (\bibinfo{year}{2018}) \bibinfo{pages}{944--954}.
\bibitem[{Geers and Hunter(2002)}]{gh02}
\bibinfo{author}{T.~Geers}, \bibinfo{author}{K.~Hunter},
\newblock \bibinfo{title}{An integrated wave-effects model for an underwater
  explosion bubble},
\newblock \bibinfo{journal}{J. Acoust. Soc. Am.} \bibinfo{volume}{111}
  (\bibinfo{year}{2002}) \bibinfo{pages}{1584}.
\bibitem[{Zhang et~al.(2023)Zhang, Li, Cui, Li, and Liu}]{unified2023}
\bibinfo{author}{A.-M. Zhang}, \bibinfo{author}{S.-M. Li},
  \bibinfo{author}{P.~Cui}, \bibinfo{author}{S.~Li}, \bibinfo{author}{Y.-L.
  Liu},
\newblock \bibinfo{title}{A unified theory for bubble dynamics},
\newblock \bibinfo{journal}{Phys. Fluids} \bibinfo{volume}{35}
  (\bibinfo{year}{2023}) \bibinfo{pages}{033323}.
\bibitem[{Zhong et~al.(2020)Zhong, Eshraghi, Vlachos, Dabiri, and
  Ardekani}]{Zhong2020}
\bibinfo{author}{X.~Zhong}, \bibinfo{author}{J.~Eshraghi},
  \bibinfo{author}{P.~Vlachos}, \bibinfo{author}{S.~Dabiri},
  \bibinfo{author}{A.~M. Ardekani},
\newblock \bibinfo{title}{A model for a laser-induced cavitation bubble},
\newblock \bibinfo{journal}{Int. J. Multiphase Flow} \bibinfo{volume}{132}
  (\bibinfo{year}{2020}) \bibinfo{pages}{103433}.
\bibitem[{Han et~al.(2023)Han, Chen, and Guo}]{Rui2023}
\bibinfo{author}{R.~Han}, \bibinfo{author}{J.~Chen}, \bibinfo{author}{T.~Guo},
\newblock \bibinfo{title}{A modified phase-transition model for
  multi-oscillations of spark-generated bubbles},
\newblock \bibinfo{journal}{Inventions} \bibinfo{volume}{8}
  (\bibinfo{year}{2023}) \bibinfo{pages}{131}.
\bibitem[{Zeng et~al.(2018)Zeng, Gonzalez-Avila, Dijkink, Koukouvinis,
  Gavaises, and Ohl}]{Qingyun2018}
\bibinfo{author}{Q.~Zeng}, \bibinfo{author}{S.~Gonzalez-Avila},
  \bibinfo{author}{R.~Dijkink}, \bibinfo{author}{P.~Koukouvinis},
  \bibinfo{author}{M.~Gavaises}, \bibinfo{author}{C.-D. Ohl},
\newblock \bibinfo{title}{Wall shear stress from jetting cavitation bubbles},
\newblock \bibinfo{journal}{J. Fluid Mech.} \bibinfo{volume}{846}
  (\bibinfo{year}{2018}) \bibinfo{pages}{341--355}.
\bibitem[{Cerbus et~al.(2022)Cerbus, Chraibi, Tondusson, Petit, Soto,
  Devillard, Delville, and Kellay}]{cerbus2022experimental}
\bibinfo{author}{R.~Cerbus}, \bibinfo{author}{H.~Chraibi},
  \bibinfo{author}{M.~Tondusson}, \bibinfo{author}{S.~Petit},
  \bibinfo{author}{D.~Soto}, \bibinfo{author}{R.~Devillard},
  \bibinfo{author}{J.-P. Delville}, \bibinfo{author}{H.~Kellay},
\newblock \bibinfo{title}{Experimental and numerical study of laser-induced
  secondary jetting},
\newblock \bibinfo{journal}{J. Fluid Mech.} \bibinfo{volume}{934}
  (\bibinfo{year}{2022}) \bibinfo{pages}{A14}.
\bibitem[{Fan et~al.(2024)Fan, Bussmann, Reuter, Bao, Adami, Gordillo, Adams,
  and Ohl}]{PhysRevLett.132.104004}
\bibinfo{author}{Y.~Fan}, \bibinfo{author}{A.~Bussmann},
  \bibinfo{author}{F.~Reuter}, \bibinfo{author}{H.~Bao},
  \bibinfo{author}{S.~Adami}, \bibinfo{author}{J.~Gordillo},
  \bibinfo{author}{N.~Adams}, \bibinfo{author}{C.-D. Ohl},
\newblock \bibinfo{title}{Amplification of supersonic microjets by resonant
  inertial cavitation-bubble pair},
\newblock \bibinfo{journal}{Phys. Rev. Lett.} \bibinfo{volume}{132}
  (\bibinfo{year}{2024}) \bibinfo{pages}{104004}.
\bibitem[{Gallo et~al.(2023)Gallo, Magaletti, Georgoulas, Marengo, De~Coninck,
  and Casciola}]{Gallo2023}
\bibinfo{author}{M.~Gallo}, \bibinfo{author}{F.~Magaletti},
  \bibinfo{author}{A.~Georgoulas}, \bibinfo{author}{M.~Marengo},
  \bibinfo{author}{J.~De~Coninck}, \bibinfo{author}{C.~M. Casciola},
\newblock \bibinfo{title}{A nanoscale view of the origin of boiling and its
  dynamics},
\newblock \bibinfo{journal}{Nat. Commun.} \bibinfo{volume}{14}
  (\bibinfo{year}{2023}) \bibinfo{pages}{6428}.
\bibitem[{Abbondanza et~al.(2023)Abbondanza, Gallo, and
  Casciola}]{Abbondanza2023}
\bibinfo{author}{D.~Abbondanza}, \bibinfo{author}{M.~Gallo},
  \bibinfo{author}{C.~M. Casciola},
\newblock \bibinfo{title}{Diffuse interface modeling of laser-induced
  nano-/micro-cavitation bubbles},
\newblock \bibinfo{journal}{Phys. Fluids} \bibinfo{volume}{35}
  (\bibinfo{year}{2023}) \bibinfo{pages}{022113}.
\bibitem[{Fuster et~al.(2010)Fuster, Hauke, and Dopazo}]{Fuster2010}
\bibinfo{author}{D.~Fuster}, \bibinfo{author}{G.~Hauke},
  \bibinfo{author}{C.~Dopazo},
\newblock \bibinfo{title}{Influence of the accommodation coefficient on
  nonlinear bubble oscillations},
\newblock \bibinfo{journal}{J. Acoust. Soc. Am.} \bibinfo{volume}{128}
  (\bibinfo{year}{2010}) \bibinfo{pages}{5--10}.
\bibitem[{Yasui(2018)}]{Yasuibook18}
\bibinfo{author}{K.~Yasui}, \bibinfo{title}{Acoustic Cavitation and Bubble
  Dynamics}, SpringerBriefs in Molecular Science, \bibinfo{publisher}{Springer
  Cham}, \bibinfo{year}{2018}.
\bibitem[{Hauke et~al.(2007)Hauke, Fuster, and Dopazo}]{Hauke2007}
\bibinfo{author}{G.~Hauke}, \bibinfo{author}{D.~Fuster},
  \bibinfo{author}{C.~Dopazo},
\newblock \bibinfo{title}{Dynamics of a single cavitating and reacting bubble},
\newblock \bibinfo{journal}{Phys. Rev. E} \bibinfo{volume}{75}
  (\bibinfo{year}{2007}) \bibinfo{pages}{066310}.
\bibitem[{Tian et~al.(2022)Tian, Zhang, Yin, Lv, and Zhu}]{Tian2022}
\bibinfo{author}{L.~Tian}, \bibinfo{author}{Y.~Zhang},
  \bibinfo{author}{J.~Yin}, \bibinfo{author}{L.~Lv}, \bibinfo{author}{J.~Zhu},
\newblock \bibinfo{title}{A simplified model for the gas-vapor bubble
  dynamics},
\newblock \bibinfo{journal}{J. Acoust. Soc. Am.} \bibinfo{volume}{152}
  (\bibinfo{year}{2022}) \bibinfo{pages}{2117--2127}.
\bibitem[{Seo et~al.(2010)Seo, Lele, and Tryggvason}]{Seo2010}
\bibinfo{author}{J.~H. Seo}, \bibinfo{author}{S.~K. Lele},
  \bibinfo{author}{G.~Tryggvason},
\newblock \bibinfo{title}{Investigation and modeling of bubble-bubble
  interaction effect in homogeneous bubbly flows},
\newblock \bibinfo{journal}{Phys. Fluids} \bibinfo{volume}{22}
  (\bibinfo{year}{2010}) \bibinfo{pages}{063302}.
\bibitem[{Brenner et~al.(2002)Brenner, Hilgenfeldt, and Lohse}]{bhl02}
\bibinfo{author}{M.~Brenner}, \bibinfo{author}{S.~Hilgenfeldt},
  \bibinfo{author}{D.~Lohse},
\newblock \bibinfo{title}{Single-bubble sonoluminescence},
\newblock \bibinfo{journal}{Rev. Mod. Phys.} \bibinfo{volume}{74}
  (\bibinfo{year}{2002}) \bibinfo{pages}{425}.
\bibitem[{Kyuichi(2021)}]{Yasui2021a}
\bibinfo{author}{Y.~Kyuichi},
\newblock \bibinfo{title}{Multibubble sonoluminescence from a theoretical
  perspective},
\newblock \bibinfo{journal}{Molecules} \bibinfo{volume}{26}
  (\bibinfo{year}{2021}) \bibinfo{pages}{4624}.
\bibitem[{Yasui(1997)}]{RN1317}
\bibinfo{author}{K.~Yasui},
\newblock \bibinfo{title}{Alternative model of single-bubble sonoluminescence},
\newblock \bibinfo{journal}{Phys. Rev. E} \bibinfo{volume}{56}
  (\bibinfo{year}{1997}) \bibinfo{pages}{6750--6760}.
\bibitem[{Schrage(1953)}]{Schrage1953}
\bibinfo{author}{R.~W. Schrage}, \bibinfo{title}{A Theoretical Study of
  Interphase Mass Transfer}, \bibinfo{publisher}{Columbia University Press},
  \bibinfo{year}{1953}.
\bibitem[{Akhatov et~al.(2001)Akhatov, Lindau, Topolnikov, Mettin, Vakhitova,
  and Lauterborn}]{Akhatov2001}
\bibinfo{author}{I.~Akhatov}, \bibinfo{author}{O.~Lindau},
  \bibinfo{author}{A.~Topolnikov}, \bibinfo{author}{R.~Mettin},
  \bibinfo{author}{N.~Vakhitova}, \bibinfo{author}{W.~Lauterborn},
\newblock \bibinfo{title}{Collapse and rebound of a laser-induced cavitation
  bubble},
\newblock \bibinfo{journal}{Phys. Fluids} \bibinfo{volume}{13}
  (\bibinfo{year}{2001}) \bibinfo{pages}{2805--2819}.
\bibitem[{Yasui(1998)}]{Yasui1998}
\bibinfo{author}{K.~Yasui},
\newblock \bibinfo{title}{Effect of non-equilibrium evaporation and
  condensation on bubble dynamics near the sonoluminescence threshold},
\newblock \bibinfo{journal}{Ultrasonics} \bibinfo{volume}{36}
  (\bibinfo{year}{1998}) \bibinfo{pages}{575--580}.
\bibitem[{de~Graaf et~al.(2014)de~Graaf, Penesis, and Brandner}]{RN1335}
\bibinfo{author}{K.~L. de~Graaf}, \bibinfo{author}{I.~Penesis},
  \bibinfo{author}{P.~A. Brandner},
\newblock \bibinfo{title}{Modelling of seismic airgun bubble dynamics and
  pressure field using the gilmore equation with additional damping factors},
\newblock \bibinfo{journal}{Ocean Eng.} \bibinfo{volume}{76}
  (\bibinfo{year}{2014}) \bibinfo{pages}{32--39}.
\bibitem[{Li et~al.(2020)Li, der Meer, Zhang, Prosperetti, and Lohse}]{shuai20}
\bibinfo{author}{S.~Li}, \bibinfo{author}{D.~V. der Meer},
  \bibinfo{author}{A.-M. Zhang}, \bibinfo{author}{A.~Prosperetti},
  \bibinfo{author}{D.~Lohse},
\newblock \bibinfo{title}{Modelling large scale airgun-bubble dynamics with
  highly non-spherical features},
\newblock \bibinfo{journal}{Int. J. Multiphase Flow} \bibinfo{volume}{122}
  (\bibinfo{year}{2020}) \bibinfo{pages}{103143}.
\bibitem[{Yasui(1999)}]{Yasui1999}
\bibinfo{author}{K.~Yasui},
\newblock \bibinfo{title}{Single-bubble and multibubble sonoluminescence},
\newblock \bibinfo{journal}{Phys. Rev. Lett.} \bibinfo{volume}{83}
  (\bibinfo{year}{1999}) \bibinfo{pages}{4297}.
\bibitem[{Yasui et~al.(2016)Yasui, Tuziuti, and Kanematsu}]{Yasui2016}
\bibinfo{author}{K.~Yasui}, \bibinfo{author}{T.~Tuziuti},
  \bibinfo{author}{W.~Kanematsu},
\newblock \bibinfo{title}{Extreme conditions in a dissolving air nanobubble},
\newblock \bibinfo{journal}{Phys. Rev. E} \bibinfo{volume}{94}
  (\bibinfo{year}{2016}) \bibinfo{pages}{013106}.
\bibitem[{Kogan(1969)}]{RN1341}
\bibinfo{author}{M.~N. Kogan}, \bibinfo{title}{Rarefied Gas Dynamics},
  \bibinfo{publisher}{Plenum}, \bibinfo{year}{1969}.
\bibitem[{Yasui(1995)}]{RN1315}
\bibinfo{author}{K.~Yasui},
\newblock \bibinfo{title}{Effects of thermal conduction on bubble dynamics near
  the sonoluminescence threshold},
\newblock \bibinfo{journal}{J. Acoust. Soc. Am.} \bibinfo{volume}{98}
  (\bibinfo{year}{1995}) \bibinfo{pages}{2772--2782}.
\bibitem[{Dai et~al.(2024)Dai, Zhu, Wang, Chu, and Wang}]{RN1370}
\bibinfo{author}{Z.~Dai}, \bibinfo{author}{J.~Zhu}, \bibinfo{author}{Z.~Wang},
  \bibinfo{author}{S.~Chu}, \bibinfo{author}{Y.~Wang},
\newblock \bibinfo{title}{Adaptive thermodynamic consistency control via
  interface thickness in pseudopotential lattice boltzmann method across wide
  temperature ranges},
\newblock \bibinfo{journal}{Phys. Fluids} \bibinfo{volume}{36}
  (\bibinfo{year}{2024}) \bibinfo{pages}{033349}.
\bibitem[{Yasui(1996)}]{RN1313}
\bibinfo{author}{K.~Yasui},
\newblock \bibinfo{title}{Variation of liquid temperature at bubble wall near
  the sonoluminescence threshold},
\newblock \bibinfo{journal}{J. Phys. Soc. Jpn.} \bibinfo{volume}{65}
  (\bibinfo{year}{1996}) \bibinfo{pages}{2830--2840}.
\bibitem[{Yasui(2001)}]{RN1339}
\bibinfo{author}{K.~Yasui},
\newblock \bibinfo{title}{Effect of liquid temperature on sonoluminescence},
\newblock \bibinfo{journal}{Phys. Rev. E} \bibinfo{volume}{64}
  (\bibinfo{year}{2001}) \bibinfo{pages}{016310}.
\bibitem[{Nagalingam et~al.(2023)Nagalingam, Raghunathan, Korede, Poelma,
  Smith, Hartkamp, Padding, and Eral}]{Nagalingam2023}
\bibinfo{author}{N.~Nagalingam}, \bibinfo{author}{A.~Raghunathan},
  \bibinfo{author}{V.~Korede}, \bibinfo{author}{C.~Poelma},
  \bibinfo{author}{C.~S. Smith}, \bibinfo{author}{R.~Hartkamp},
  \bibinfo{author}{J.~T. Padding}, \bibinfo{author}{H.~Eral},
\newblock \bibinfo{title}{Laser-induced cavitation for controlling
  crystallization from solution},
\newblock \bibinfo{journal}{Phys. Rev. Lett.} \bibinfo{volume}{131}
  (\bibinfo{year}{2023}) \bibinfo{pages}{124001}.
\bibitem[{Chen et~al.(2024)Chen, Chen, Geng, Huang, and Cao}]{Jiacheng2024}
\bibinfo{author}{J.~Chen}, \bibinfo{author}{T.~Chen},
  \bibinfo{author}{H.~Geng}, \bibinfo{author}{B.~Huang},
  \bibinfo{author}{Z.~Cao},
\newblock \bibinfo{title}{Investigation on dynamic characteristics and thermal
  effects of single cavitation bubble in liquid nitrogen},
\newblock \bibinfo{journal}{Phys. Fluids} \bibinfo{volume}{36}
  (\bibinfo{year}{2024}) \bibinfo{pages}{023325}.
\bibitem[{Li et~al.(2024)Li, Zhao, Zhang, and Han}]{RN1269}
\bibinfo{author}{S.~Li}, \bibinfo{author}{Z.~Zhao}, \bibinfo{author}{A.-M.
  Zhang}, \bibinfo{author}{R.~Han},
\newblock \bibinfo{title}{Cavitation bubble dynamics inside a droplet suspended
  in a different host fluid},
\newblock \bibinfo{journal}{J. Fluid Mech.} \bibinfo{volume}{979}
  (\bibinfo{year}{2024}) \bibinfo{pages}{A47}.
\bibitem[{Liang et~al.(2022)Liang, Linz, Freidank, Paltauf, and
  Vogel}]{Liang2022}
\bibinfo{author}{X.~X. Liang}, \bibinfo{author}{N.~Linz},
  \bibinfo{author}{S.~Freidank}, \bibinfo{author}{G.~Paltauf},
  \bibinfo{author}{A.~Vogel},
\newblock \bibinfo{title}{Comprehensive analysis of spherical bubble
  oscillations and shock wave emission in laser-induced cavitation},
\newblock \bibinfo{journal}{J. Fluid Mech.} \bibinfo{volume}{940}
  (\bibinfo{year}{2022}) \bibinfo{pages}{A5}.
\bibitem[{Han et~al.(2022)Han, Zhang, Tan, and Li}]{Hanrui2022}
\bibinfo{author}{R.~Han}, \bibinfo{author}{A.-M. Zhang},
  \bibinfo{author}{S.~Tan}, \bibinfo{author}{S.~Li},
\newblock \bibinfo{title}{Interaction of cavitation bubbles with the interface
  of two immiscible fluids on multiple time scales},
\newblock \bibinfo{journal}{J. Fluid Mech.} \bibinfo{volume}{932}
  (\bibinfo{year}{2022}) \bibinfo{pages}{A8}.
\bibitem[{Wang(2013)}]{w13}
\bibinfo{author}{Q.~Wang},
\newblock \bibinfo{title}{Non-spherical bubble dynamics of underwater
  explosions in a compressible fluid},
\newblock \bibinfo{journal}{Phys. Fluids} \bibinfo{volume}{25}
  (\bibinfo{year}{2013}) \bibinfo{pages}{072104}.
\bibitem[{Wang(2016)}]{Qianxi2016}
\bibinfo{author}{Q.~Wang},
\newblock \bibinfo{title}{Local energy of a bubble system and its loss due to
  acoustic radiation},
\newblock \bibinfo{journal}{J. Fluid Mech.} \bibinfo{volume}{797}
  (\bibinfo{year}{2016}) \bibinfo{pages}{201--230}.

\end{thebibliography}



\end{document}